%% file: main.tex
\documentclass[preprint,12pt]{elsarticle}

\usepackage{amssymb}

\usepackage{xcolor}
\usepackage{soul}
\usepackage[colorlinks]{hyperref}
\usepackage{booktabs}
\usepackage{multirow}
\usepackage{colortbl}
\usepackage{tikz}
\usetikzlibrary{calc,positioning}
\usepackage{array}
\usepackage{tabularx}

\definecolor{OCircle}{HTML}{AB73FF}
\definecolor{CCircle}{HTML}{B80833}

\newcommand{\ovcircle}[1]{
\begin{tikzpicture}[enum/.style={circle,draw=gray,very
thin,fill=OCircle,text=white,inner sep=1pt},baseline=-3pt]
  \node (node 1) at (0,0) [enum] {\scriptsize #1};
 \end{tikzpicture}}
 
 \newcommand{\ccircle}[1]{
\begin{tikzpicture}[enum/.style={circle,draw=gray,very
thin,fill=CCircle,text=white,inner sep=1pt},baseline=-3pt]
  \node (node 1) at (0,0) [enum] {\scriptsize #1};
 \end{tikzpicture}}

\newcommand{\yellowdot}{\tikz\fill[olive] (0,0) circle (.5ex);}
\newcommand{\orangedot}{\tikz\fill[orange] (0,0) circle (.5ex);}

\newcommand{\mysec}[1]{\smallskip \noindent \textbf{#1:}}

\begin{document}

\begin{frontmatter}



\title{``My toxic trait is thinking I’ll remember this'':\\ gaps in the learner experience of video tutorials for feature-rich software}


\affiliation[inst1]{organization={Microsoft Research},
            city={Cambridge},
            country={United Kingdom}}

\affiliation[inst2]{organization={University of Cambridge},
            city={Cambridge},
            country={United Kingdom}}
            
\affiliation[inst3]{organization={University College London},
            city={London},
            country={United Kingdom}}

\author[inst1]{Ian Drosos}
\author[inst1,inst2,inst3]{Advait Sarkar}
\author[inst1]{Andrew D. Gordon}

\begin{abstract}
Video tutorials are a popular medium for informal and formal learning. However, when learners attempt to view and follow along with these tutorials, they encounter what we call gaps, that is, issues that can prevent learning. We examine the gaps encountered by users of video tutorials for feature-rich software, such as spreadsheets. We develop a theory and taxonomy of such gaps, identifying how they act as barriers to learning, by collecting and analyzing 360 viewer comments from 90 Microsoft Excel video tutorials published by 43 creators across YouTube, TikTok, and Instagram. We conducted contextual interviews with 8 highly influential tutorial creators to investigate the gaps their viewers experience and how they address them. Further, we obtain insights into their creative process and frustrations when creating video tutorials. Finally, we present creators with two designs that aim to address gaps identified in the comment analysis for feedback and alternative design ideas.
\end{abstract}




\end{frontmatter}


\newpage
\section{Introduction}
\begin{figure}
  \includegraphics[width=\textwidth]{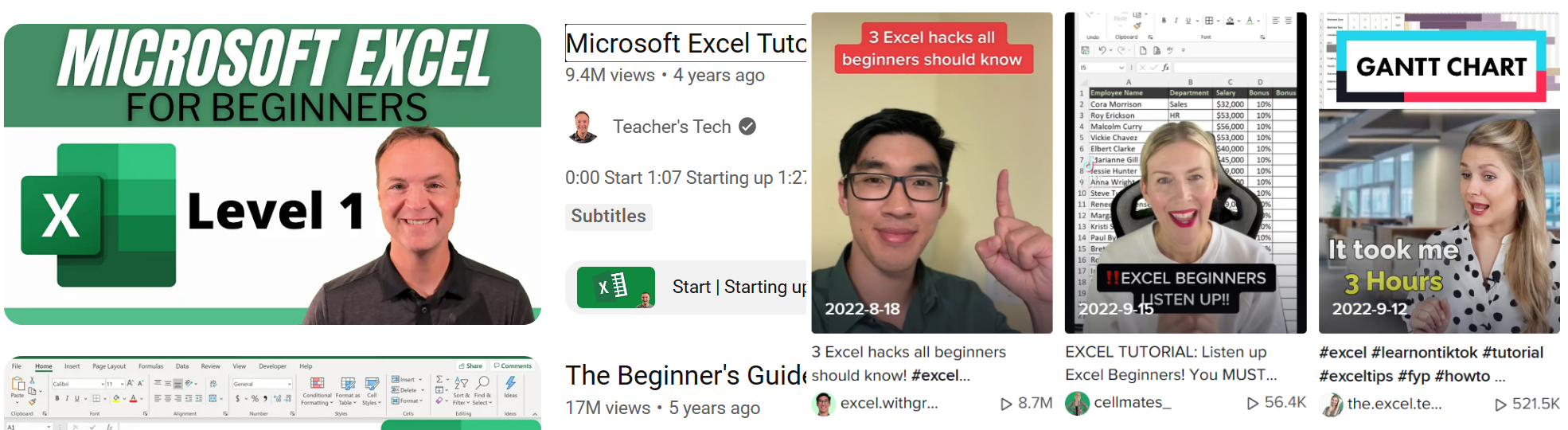}
  \caption{Example of Excel video tutorials on YouTube (left) and TikTok (right): we find gaps that occur when learners use such videos.}
  \label{fig:teaser}
\end{figure}

\label{sec:intro}
Video tutorials are a popular medium for informal and formal learning at scale about concepts and tools such as art, coding, and working with spreadsheets \cite{kim2014crowdsourcing,kiani2019beyond,end-user-encounters-with-lambda-22}. However, several issues may prevent learners from using these tutorials successfully. These represent \emph{gaps} between the content creator and their video tutorial and a viewer's own context. We define a gap to be \emph{any barrier a user faces in successfully learning and applying the knowledge content of the tutorial} (this definition is explained in greater detail in Section~\ref{sec:background}).


To better understand the nature of gaps in the learner experience of video tutorials, we investigated viewer reactions to video tutorials, and their difficulties in watching and learning from them, through a content analysis of comments on video tutorials for the Microsoft Excel spreadsheet software. 

Based on this analysis, we developed two design prototypes aimed at content creators to help them better anticipate and address learner gaps during the video tutorial creation process. We then interviewed influential content creators who create video tutorials for Excel about their creative process and frustrations with this process, and about learner gaps they have seen their viewers struggle with and how they address them. We also collected their feedback on the design prototypes.


\subsection{Excel: an example of feature-rich software}
Why study Microsoft Excel in particular? 
The challenges associated with learning from video tutorials extend to all software, and indeed many of our findings and implications for design are not specific to spreadsheets; one approach to tackling our research questions may have been to study a wide variety of applications to synthesize common issues. However, given the enormous diversity of software and its attendant tutorial landscape, to manage the scope and complexity of this investigation and data analysis, we chose to focus on one application, an approach commonly taken by studies of software help systems (Section~\ref{sec:background}).


Video tutorials for Excel began widely circulating in the mid-2000s.
The earliest video on the \emph{MrExcel.com} YouTube channel, for example, was posted on 23 July 2006.\footnote{\url{https://www.youtube.com/watch?v=cb-8wNd0Srk}}
The channel is part of the tutorial website \url{MrExcel.com} that was founded in 1998.\footnote{\url{https://www.mrexcel.com/about-mrexcel/}}
By 2014, video tutorials on YouTube (although not specifically on Excel) had become the subject of academic research \cite{kim2014crowdsourcing}. 
By 2015, there was an Excel MOOC \emph{EX101x: Data Analysis: Take it to the Max()}\footnote{\url{https://web.archive.org/web/20160415215403/https://www.edx.org/course/data-analysis-take-it-max-delftx-ex101x-0\#!}}, which ran twice and reached almost 60,000 students \cite{DBLP:conf/wcre/RoyHAWD16}.
%
%
There is evidence that video tutorials for Excel grew dramatically in popularity in the second half of the 2010s.
%
An interview study of Excel learners, conducted in 2016, makes no mention of video tutorials or YouTube \cite{sarkar2018spreadsheetlearning}, but by 2020, at least, there were many video tutorials with hundreds of thousands of views, and with long comment sections suitable for studying perceptions of the software \cite{end-user-encounters-with-lambda-22}.
%
Excel video tutorials can now be found on YouTube, Instagram, and TikTok, which are major social media platforms for consuming video content with over a billion users each \cite{datareport}.
Video tutorials have thus been available for Excel since at least 2006, and in the years following 2016 they became a popular mechanism for self-instruction.

%
%

Excel is a suitable choice for study because it typifies a kind of feature-rich software\footnote{If any evidence is needed that Excel is feature-rich, take the book \emph{Microsoft Excel Data Analysis and Business Modeling} by Wayne L. Winston, a comprehensive guide to the software. Its first edition, published in 2007, has 624 pages; the most recent edition, the seventh, published in 2021, has 1168 pages.} 
that is used intensively by many professionals, is frequently updated, and where there is a constant need and motivation to learn for large portions of its user base.
As such, it falls into the same category of feature-rich software as computer-aided design (CAD) tools such as AutoCAD, or graphics packages such as Adobe Photoshop, termed ``praxisware'' \cite{sarkar2023should}. This results in a rich and varied ecosystem of video tutorials, and a diverse set of users who engage with this ecosystem for many different purposes.
Moreover, while many previous studies of video tutorials and software help systems used CAD or graphics applications as their domains of study, spreadsheets are relatively under-explored (Section~\ref{sec:background}). The contrast in the typical use cases of spreadsheets, and the learning challenges of end-user programmers \cite{ko2004six,sarkar2018spreadsheetlearning}, versus those of CAD or graphics applications, makes the contributions of our study a novel and valuable addition to our knowledge of end-user learning of praxisware.

\subsection{Overview of findings and contributions}

We report, to our knowledge, the first study of gaps between Excel video tutorials and learner practice, supplemented by creator interviews who provide insights into the creation process. 

%

Our analysis of comments on video tutorials revealed 13 gaps (found in Table \ref{tbl:gaps}) 
that are barriers to learning (Section \ref{sec:resultgaps}).
We found five categories of \emph{creator-driven gaps} (that is, barriers to learning attributable to the video creator), four categories of \emph{learner-driven gaps} (that is, attributable to the learner), and four categories of \emph{app-driven gaps}, (that is, attributable to the application).
These categories related to spreadsheet sharing, production issues, missing steps, and clarifying choices. In our interviews, creators discussed how they address user feedback reporting such gaps.

%


Our interviews with video tutorial creators shed light on their process and frustrations when creating tutorial videos. We identified five interview themes concerning the \emph{creative process} (Section~\ref{sec:resultprocess}): picking a topic, creating spreadsheets, editing and production, how media platforms affect video tutorials, and regarding the motivation to create. We identified seven interview themes concerning \emph{creative frustrations} (Section~\ref{sec:resultfrustration}): 
with Excel,
with creation tools,
sharing knowledge with learners,
making effective tutorial videos,
creating lessons,
with mistakes they make,
and finally features that ease frustrations.

We finally presented creators with two design prototypes, based on gaps identified in the analysis of comments. Creators gave feedback on how these designs would be useful to address many of the gaps we found in our comment analysis. Creators believed the interactive video tutorial system described in Design 1 would assist in the learning journeys of learners, and helped alleviate common production issues that impact learners. Creators also thought gap detection and intervention aspects of Design 2 would be useful for closing many of the gaps that learners face and help creators make more effective tutorials. Finally, creators spoke about how these designs would need to go farther to address many of the challenges they see with video tutorial creation and consumption.







In sum, the contributions of this paper are:
\begin{itemize}
    \item An analysis of 360 comments, from 90 video tutorials, about gaps encountered by learners when watching and learning from the tutorial. We detail 13 gaps that can occur when learners attempt to practice what they see on tutorials.
    \item Contextual interviews with eight creators of video tutorials for Excel, comprising:
    \begin{itemize}
        \item an investigation of their frustrations with creating video tutorials, learner gaps and how they address them.
        \item a design probe of two designs aimed at addressing gaps identified in the comment analysis, to discover design implications for helping creators produce more effective video tutorials.
    \end{itemize}
\end{itemize}

Though our focus was on Excel video tutorials and their creators,
our findings have implications (discussed in Section~\ref{sec:disc}) beyond spreadsheets and likely to apply to other feature-rich software. We find creators need support for creating effective video tutorials that directly interact with the application and consider platform restrictions. We also find that creators need assistance in addressing gaps in their videos and or in interactions with learners in comment sections. Additionally, we find that learners need better help selecting the next tutorial to view and finding tutorials that closely align with their own context, skill or knowledge level, and learning goals.

\subsection{Structure of the paper and study overview}
\label{sec:methods}
%


\begin{figure}
    \centering
\includegraphics[width=\textwidth]{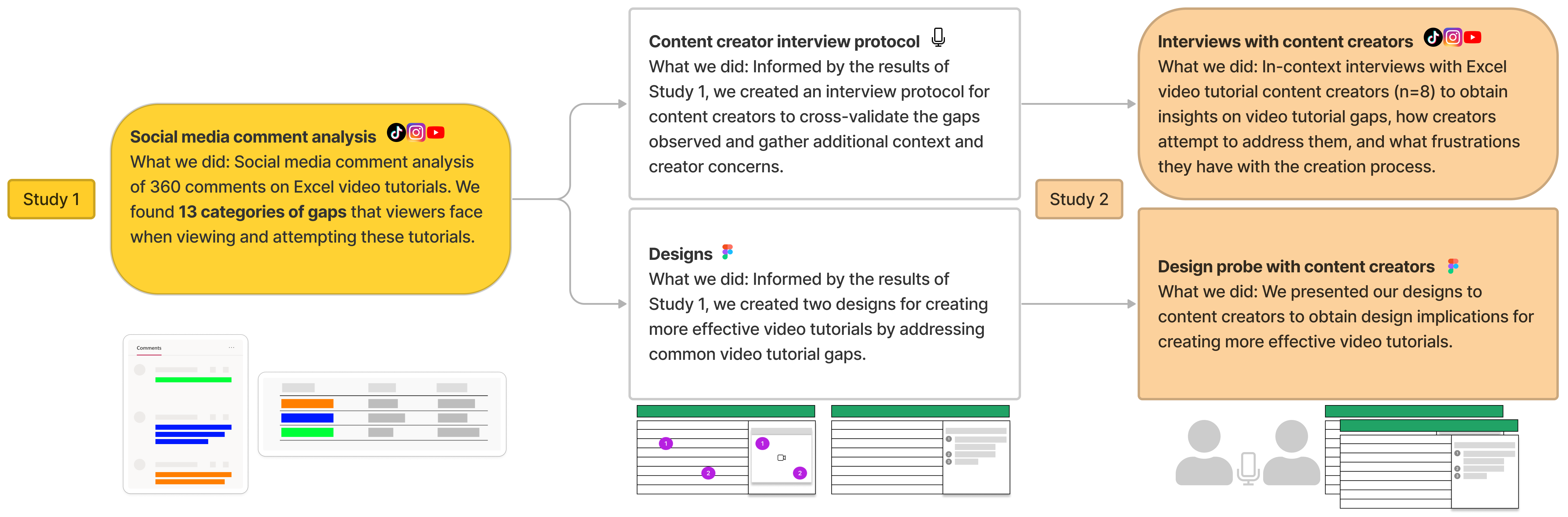}
    \caption{Diagram of the work presented in this paper. Study 1 is a social media comment analysis to discover categories of gaps that viewers face when using video tutorials (Sections \ref{sec:commentmethods} and \ref{sec:resultgaps}). Based on Study 1, we created two designs that attempted to address many of the gaps we had found (Section \ref{sec:dprobemethods}). We also created an interview protocol to gain insights from content creators (Section \ref{sec:interviewmethods}). In Study 2, we interviewed content creators about these gaps, their creation workflows, and their frustrations (Section \ref{sec:results}).  Finally, we presented our designs to creators to obtain design implications for effective video tutorials (Section \ref{sec:resultdprobe}).}
    \label{fig:VSD}
\end{figure}

In Section \ref{sec:background}, we discuss what a video tutorial gap is, and review related work in improving video tutorials and software learning.

Figure \ref{fig:VSD} is an overview of our investigation. We first conducted a qualitative study of gaps experienced by video tutorial viewers, by analyzing viewer comments on video tutorials (Sections \ref{sec:commentmethods} and \ref{sec:resultgaps}). Based on this analysis, we created two designs aimed at addressing several common gaps (Section \ref{sec:dprobemethods}), and developed an interview protocol for content creators (Section \ref{sec:interviewmethods}), to triangulate the experiences of viewers with those of creators. 

We then interviewed content creators who publish Excel tutorials on various platforms to gain their insights on these gaps, how they address them, and on their creative process (Section \ref{sec:results}). Through a probe using the two designs, we elicited design ideas from content creators for addressing gaps in the video tutorial experience (Section \ref{sec:resultdprobe}).

\input{tbl_findings}

Table \ref{tbl:findings} lists the themes we found over the course of our investigation, including video tutorial gaps, creator processes and frustrations, and design implications for future learning tools. Table \ref{tbl:findings} shows how each component of our investigation (comment analysis, creator interviews, and design probe) informed our understanding of each theme. The themes will be described in detail in later sections.

We discuss design implications and study limitations in Section \ref{sec:disc}.
Section~\ref{sec:conclusion} concludes.

\section{Background}
\label{sec:background}
We propose to identify gaps in video tutorials for learning feature-rich software. This raises the question: what is a gap?


In theory, a gap can be defined as any departure from a hypothetical ``ideal tutorial'' experience that, for any given user, contains no gaps. The problem with a notion of ideal tutorial is that it is trivially satisfied by a detailed, step-by-step solution of the user's particular task at hand. Such a tutorial would solve the user's immediate problem, but may not give the user lasting knowledge or confidence in tackling related problems in the future, i.e., it is ``giving the user the fish'' rather than teaching them how to fish, a problem related to information retrieval's ``dilemma of the direct answer'' \cite{potthast2021dilemma}. 

In some cases, gaps can be good; the synthesis and interpretation required to bridge the gap results in a more durable form of constructivist learning \cite{sarkar2016constructivist}. It can be questioned whether these are even gaps, or whether they are an intended feature of the learning process, of the ideal tutorial.

Some gaps can be self-diagnosed by the learner, others cannot (because they require expertise to identify). In the next section, we discuss research that has found how the proliferation of video tutorials results in an exploration-exploitation tradeoff; the problem of locating the ``ideal'' tutorial in a vast landscape of similar online tutorials, versus working with the tutorials that appear early on during the search and information foraging \cite{srinivasa2016foraging} process to make it work for you.

Rather than impose a rigid definition of ``gap'', we allowed the definition to emerge from our qualitative analysis of real users engaging with video tutorials, and content creators struggling to impart lasting knowledge. Consequently, we adopted the definition of a gap as \emph{any barrier a user faces in successfully learning and applying the knowledge content of the tutorial}. 




\subsection{Improving video tutorials}
Grossman et al. characterize the many challenges of software learnability, particularly with feature-rich software such as spreadsheets \cite{grossman2009survey}. In parallel, it was shown that graphic or animated visuals make software help instructions much more effective \cite{harrison1995comparison,chi2012mixt}. These findings motivated research on the use of videos for software learning.

Finding the right video tutorial to watch can be a challenge. Fourney et al. explore automatically building ``query-feature'' graphs which map associations from natural language statements of user goals and relevant software features \cite{fourney2011query}. Others have developed multimodal search features where a query can be authored through, for example, pointing at a UI element (rather than having to know the formal name of the feature) \cite{fraser2020remap}, or automatically contextualizing a tutorial search with terms drawn from the user's current application and activity \cite{fraser2019replay}. Because of the wide availability of many related tutorials, participants often prefer to search until they find a video that appears to match their current goal closely, rather than attempt to work through a tutorial that may not satisfy their requirements. Yet, it is difficult for non-experts to judge whether a video will suit their needs until they have substantively engaged with it. This attention investment tradeoff is termed the ``paradox of videos'' by Kiani et al. \cite{kiani2019beyond}.

Despite the paradox (or perhaps to take advantage of it), many proposals aim to greatly increase the speed and quantity of tutorial materials. Multiple projects have explored automated or partially-automated generation of documentation and tutorials from demonstrations by the user \cite{bergman2005docwizards, leshed2008coscripter, grabler2009generating, lafreniere2014investigating}. Others have demonstrated crowdsourcing approaches to personalized tutorial generation \cite{whitehill2017crowdsourcing}.

Upon finding the right video, the viewer has many challenges in successfully exploiting it as a learning resource. Guo et al. characterize the important distinction of tutorial videos: they present \emph{procedural} (how-to) knowledge, as opposed to lecture videos, which present conceptual knowledge \cite{guo2014video}. As such, rather than consumed once linearly, tutorial videos are intrinsically composed of substeps, and users actively watch these videos by navigating back and forth \cite{kiani2019beyond}, using speed controls, and repeated watching, while following along themselves in their own application \cite{chi2012mixt}. 

Hence, a lot of research focused on improving affordances for navigating and following tutorial videos. Linking videos with applications can ameliorate the cumbersome nature of ``pause and play'' watching \cite{pongnumkul2011pause}. For videos not already organized into neat substeps, these can be automatically generated by analyzing the video content \cite{pavel2014video}, or can be `learnersourced' from viewers themselves \cite{weir2015learnersourcing, kim2014crowdsourcing}. Video navigation can be enriched to support patterns such as rewatching, textual/visual search, and skimming \cite{kim2014data,kim2014understanding}.

\subsection{Improving software learning}
Many results from research into software help systems, not necessarily involving video, are germane to the challenges of video tutorials. Users are reticent to read documentation \cite{rettig1991nobody} and may prefer to learn by exploration \cite{rieman1996field}, or from more experienced colleagues \cite{sarkar2018spreadsheetlearning}, though research has found that many user frustrations could be addressed through training \cite{lazar2006workplace}. This is referred to as the `paradox of the active user' \cite{carroll1987paradox}. In response, researchers have proposed `minimal' learning interfaces, such as tooltips \cite{farkas1993role,huang2007graphstract}. It is challenging to create and distribute how-to documentation across an organization, a problem tackled in many ways, including automatic generation from demonstrations \cite{bergman2005docwizards,leshed2008coscripter,grabler2009generating, lau2004sheepdog}. It is challenging to follow tutorial steps when the user interface (UI) is complex and relevant UI elements are hard to locate; this can be ameliorated by automatically highlighting only the relevant portions of the UI \cite{kelleher2005stencils}. Feature rich software can have progressive disclosure, so that advanced features only appear after a certain level of expertise has been reached \cite{carroll1984training, mcgrenere2002evaluation, mcgrenere2007field}, or restructure itself according to the task \cite{lafreniere2014task}. Documentation itself may be automatically tailored to user expertise \cite{andrade2008expressing}. Help systems can be passive (i.e., invoked by the user) or proactively provide help at appropriate moments, though the evidence for effectiveness and user acceptance of proactive help is mixed \cite{van2006paradox, bar2019good, xiao2004empirical}. Long-term recall is also a challenge, though relatively under-explored. Grossman et al. found that interventions such as forced use and audio reinforcement can improve recall of keyboard shortcuts \cite{grossman2007strategies}.

In summary, previous work has aimed at improving the experience of finding, navigating, and following video tutorials. It has identified differences in user preferences for software learning and proposed a range of approaches, from crowdsourcing and automatic generation, to software customization and proactive intelligent assistance. 

However, the full range of potential barriers to learning and applying the knowledge content of step-by-step tutorial videos for feature-rich software, as experienced by both users and content creators, is not well-explored. This is an important space as it sits at the intersection of concerns faced daily by millions of users of feature-rich software, and the creators who aim to satisfy their needs. Identifying these issues and their relative impact on learner experiences is the first step to developing new design interventions to promote learning of feature-rich software through video tutorials. It is this gap that we aim to fill.

\section{Study 1 method: social media comment analysis}
\label{sec:commentmethods}


We were motivated to understand the gaps experienced by a wide range of users encountering video tutorials. We were also interested in how experiences of gaps in video tutorials can develop and evolve in discussion amongst end-users and content creators, something that is difficult to elicit through controlled experiments or in interviews. Platforms such as YouTube, Instagram, and TikTok create ecosystems within which viewer experiences are documented, and studying them ``can yield insights into qualitative research topics, with results comparable to and sometimes surpassing traditional qualitative research techniques'' \cite{barik2015heart}. Sarkar et al. applied this method to study the LAMBDA feature in Microsoft Excel, being the first systematic qualitative study of online communities in spreadsheet research \cite{sarkar2022lambdas}. Drawing on the experiences of those studies, we began our investigation by studying evidence of gaps experienced by viewers as reported in their online comments.

To find Excel video tutorials we searched 3 popular platforms for publishing videos: YouTube, TikTok, and Instagram. We searched between November and December 2022 using the term ``Excel Tutorial'' which returned a range of videos that included general Excel tutorials and tutorials for specific functions, features, and styles of analysis. 
We excluded videos that were not relevant (e.g., advertising courses, non-Excel tutorials) or did not have enough engagement (defined as having at least 4 comments relevant to issues viewers were having with the tutorial).


For Instagram, at the time of data collection a search for ``Excel Tutorial'' did not provide a list of videos (called ``Reels'') due to a limitation of Instagram's search. Instead, we first searched for ``Excel'' which yielded a list of content creators who make videos about Excel, and then we filtered and selected videos on these creators' pages that met the same criteria as the YouTube and TikTok videos.

We collected 360 comments, distributed evenly between YouTube, TikTok, and Instagram, from 90 videos published by 43 creators. These videos are fully listed in Appendix~\ref{apx:analyzed-videos-tables} (Tables \ref{tbl:videos}, \ref{tbl:ttvideos}, and \ref{tbl:igvideos}). 
The videos had between tens of thousands to millions of views each.
We applied inductive thematic analysis, annotating each comment and categorizing them into different types of gaps individually, followed by negotiated agreement \cite{braun2006using,campbell2013coding,mcdonald2019reliability,saldana2021coding}, where at least two researchers coded each comment, and then met to address and settle any disagreements.

\input{tbl_commentgaps.tex}







\section{Study 1 results: Finding the gaps}
\label{sec:resultgaps}

Table~\ref{tbl:gaps} lists our coded gaps and their descriptions. We identified 13 types of gaps that viewers encounter when viewing and applying video tutorials. These gaps are now detailed in turn. 


\subsection{Creator-driven gaps}
\mysec{Spreadsheet sharing}
Lack of access to the files used by the creator in the tutorial creates a gap when learners attempt to follow along with the tutorial but do not have a representative spreadsheet to do so.
We found comments that spoke to this need, for example: \emph{``in order to follow your instructions closely, I need to practice on similar work sheets in real time''} (C113). 
Comments alluded to spreadsheet downloads being useful as a way to save the learner time from manually \emph{``typing it all''} themselves (C110).
Gaps can remain even when the video creator is aware of this need and has attempted to provide files. For example, the spreadsheet might be the incorrect file, the link might be broken or unusable (\emph{``The link to the sample spreadsheet has expired.''} (C42)), or the spreadsheet in a state of partial completion that prevents effective usage by learners.


\mysec{Production issues}
Some gaps were the result of production issues. These can result from the mobile devices that many learners on TikTok and Instagram used to watch tutorials, where the small screen causes visibility issues, and videos recorded in a horizontal landscape format are difficult to view and navigate when the mobile device is held in a vertical portrait orientation. For example: \emph{``What is the function? The letters are not recognizable on the phone''} (C239). Creators mentioned that they zoomed into elements of the UI during screen recording or in edits done in post-processing to help alleviate this, which was a common request by learners when it was difficult to see the spreadsheet in the tutorial. 

Pacing was another issue, where some commenters would ask creators to \emph{``try slower video and explanation, it’s better for those who want to learn''} (C242). Some learners mentioned having to watch the video several times to understand what is happening. Music was also mentioned as a distraction, but for platforms like TikTok, music is common in videos for both education and entertainment. 

Creators also occasionally made mistakes. Comments pointed out mistakes like the keyboard shortcut used in the video was different from what the creator said in the voice-over, issues in the spreadsheet data, misselecting cells in a range for a certain formula, or even mistakes that the creator fixed but edited out in the process of doing so. The latter caused confusion in one commenter, who thought the tutorial was going to show them how to fix the issue.

\mysec{Missing steps}
Some tutorials were missing steps. For example, when a creator uses a keyboard shortcut, they may visualize the keys pressed or verbally mention them, but occasionally a creator will not inform viewers of what keys are being pressed. Sometimes a creator will skip steps to shorten the tutorial, which can be useful for short form videos on TikTok and Instagram. However, this can cause confusion when trying to follow along step-by-step. In one tutorial the creator pre-processed their data, leaving one learner to comment \emph{``You could’ve mentioned you have to have your data in a table format''}(C283). This prompts requests for \emph{``step-by-step instructions''} (C312). To us, this suggests that writing detailed step-by-step instructions might be a helpful auditing exercise for creators to make sure tutorials can be completed.

\mysec{Clarifying choices}
Commenters asked the creator to explain why they chose a particular feature or technique. Excel offers many alternative ways to accomplish a task. Commenters would often bring up their preferred alternative. For example, one commenter asked \emph{``why not use filters''} (C143), possibly trying to understand why a certain alternative was chosen. Another example involved when a tutorial was explaining a keyboard shortcut for resizing cells, a commenter suggested a mouse-centric way to perform the same task and stated that it was \emph{``simpler''} (C152) than what the creator showed. On a video that explained the VLOOKUP function, one commenter asked \emph{``why are you still using vlookup instead of xlookup''} (C175), likely since XLOOKUP is the updated function meant to replace VLOOKUP. 


Moreover, creators were also asked to clarify their choice of function parameters or values they use in a tutorial. For example, one commenter asked \emph{``why do we use number 5 after A2 (e.g., A2;5)''} (C198), which seeks to better understand how to pick parameters for certain functions, rather than questioning the motivation for selecting a certain topic or function.


\subsection{Learner-driven gaps}
\label{sec:learner-driven-gaps}

\mysec{Learning journey}
A diverse category of gaps related to the learner's journey. This can involve a learner's progression from beginner to expert, with one commenter asking the creator to \emph{``please arrange your Excel playlist from beginners to advanced level. I am confused which I should see first''} (C53). Other comments related to specific personal career goals. Comments asked which videos would help them accomplish their goals. For example, one commenter asked a creator to \emph{``please recommend more videos in reference to learning about becoming a data analyst? I started learning about the certification course fairly recently, and I am eager to learn as much as possible''} (C49). This and many comments are related to finding the right tutorial. Sometimes there were comments on videos requesting help to find a tutorial they had seen before but could no longer find it, e.g., \emph{``I am searching for the post that shows how to format a table that is copied from a webpage. The table I have is huge and will take forever to organize it''}(C289). One commenter even expressed a longing for the virtual assistant `Clippy', possibly pointing to a need for on-demand help.

\mysec{Adapting tutorials}
Learners can run into issues when applying tutorials to their own data. For example, on a tutorial about splitting 1 name column into 3, a commenter asked \emph{``when splitting names into First, Middle, Last, how do you work around names that have no middle name?''} (C95) since the video did not show how to resolve this issue. Commenters also asked questions about variations of the scenario shown in the tutorial and how the actions they took might affect what they did in the spreadsheet, for example \emph{``does the pivot table update automatically, if I change the data?''}(C99), and asked about what to do in situations where the learner has to debug or fix an error.

Learners often wish to go further than what the tutorial shows. This can be requests for more complex scenarios, for example \emph{``how do you do this if the data is in two separate Excel workbooks?''} (C237) or returning a letter grade, which goes beyond the tutorial showing how to convert a grade to pass or fail. Some learners left comments asking about other scenarios in which the taught skill could be applied. One commenter asked \emph{``how do we know it’s accurate?''} (C133) to understand ways to validate what was done in the tutorial.

Commenters asked about finding or having the creator produce alternate tutorials. These included changes to a step in the tutorial (e.g., \emph{``how can I start my first record on row 1?''} (C14)), doing the same types of things in a different application like Google Sheets or Microsoft Word, or even do the reverse of what was shown (e.g., \emph{``Similarly, how to split?''} (C327) on a tutorial about combining cells). Creators noted that these requests, particularly frequent ones, were a source of video ideas.


\mysec{Difficulties maintaining knowledge or skills} Some comments related to the learner maintaining what they learned from these tutorials, or as one commenter so eloquently put it, \emph{``My toxic trait is thinking I’ll remember this''} (C191).

Many noted the effort needed to commit knowledge to memory, especially when there are many steps. For example, \emph{``this video was only 50 minutes long but took me the best part of a day to go through it as I was making notes and doing the same thing on my PC in an attempt to remember everything''} (C68). Some watched the same video several times to help them remember the skills shown. Others noted that onscreen annotations of keyboard shortcuts used helped to remember them. One commenter asked how they were supposed to retain skills if they were not using Excel every day.



\mysec{Unknown concepts}
Learners sometimes needed explanations of concepts that the tutorial built upon. For example, on a tutorial about how to use pivot tables, one commenter said \emph{``I wish you would've explained things like what a pivot table is''} (C63).
Some learners were confused about onscreen annotations: \emph{``It would be helpful to explain what is H O I''} (C123), which was the keyboard shortcut to complete the step. Some comments requested explanation of what certain parameters in a function meant, or for a deeper explanation of what a function does, its advantages over other functions, and other situations where it applies. Creators needed to strike a balance: they cannot explain from a low level in every tutorial (Section \ref{sec:resultfrustration}).

\subsection{App-driven gaps}

\mysec{Region and language issues}
The geographic region of the user, and their preferred language, affected their Excel functionality, due to localized versions of the software. For some regions of Excel the separator between parameters is a semicolon (;) instead of a comma (,) as it is in the English versions of Excel which was noted by a commenter:  \emph{``If it doesn't work for you, try [... to] use semicolons''} (C82). Another asked about the vocabulary used by the creator for the parenthesis symbol (`('), which they called `brackets'. Finally, one commenter asked about how to convert a tutorial that used the United States dollar sign to their own country's currency, which did not have a similar currency symbol.


Language can be a gap for viewers and creators alike (Section \ref{sec:resultfrustration}), with some viewers requesting tutorials in their preferred language. Excel functions have different names in depending on the language setting, which can be an issue for multilingual learners who are watching a video in English but work in a different language. For example, one commenter requested the creator to \emph{``please include the name of the function in Spanish, too [...] a lot of us have our keyboards and systems set in Spanish''} (C247).




\mysec{Version issues}
Version issues 
led to many comments of this kind: \emph{``What version of Excel do we need to use this?''} (C325). Excel has several supported endpoints including Web, iOS, Android, Windows, Mac, and multiple versions on each platform. Many said they had issues with not having the functions being shown in the tutorial due to having an older version of Excel than the creator, e.g., \emph{``I don't have that feature, my Excel is 2016''} (C164), or using a different platform with potentially different keyboard shortcuts, e.g., \emph{``Is there another way to do it in Mac''} (C104). Further, a user might not have updated their software, meaning features may be missing even though they are on the same platform used in the tutorial.



\mysec{Configuration question}
Corporate policies may block functionality:  \emph{``most governments block unsigned macros and won’t let workers create them''} (C188). Other comments noted that the font needed in a tutorial was not in the list of choices visible in their installation of Excel, meaning the learner would first need to download, install, and enable the font in Excel. One commenter asked about if they could still follow the tutorial if they had compatibility mode enabled.


\mysec{Unexpected behavior}
Learners faced unexpected behaviors or results from following along with the tutorial. Errors can prevent completion of the tutorial (e.g., \emph{``it brings an error \#NAME?''} (C183) or other built-in errors). Some learners followed the steps in the tutorial, but there was a mismatch between their intermediate or final results and those shown. Learners sought alternative ways to get the expected behavior, e.g., \emph{``mine did not autofill down as yours did. What is the easiest way if it does not automatically do it?''} (C93). Mismatches can be the result of minor UI differences between the learner's and creator's spreadsheet. One commenter asked why the values in one of their columns were filled with `\#' characters (meaning the column was not wide enough to show the values). Finally, several comments asked about data value discrepancies with the creator's results or even external calculators, when validating the value calculated by their spreadsheet in their own implementation of the tutorial.


\section{Design probe formation}
\label{sec:dprobemethods}


We observed that despite creators' efforts to make tutorials useful and effective, several gaps remained which created barriers for user learning. This parallels findings by Ragavan et al. \cite{ragavan2021comprehension} that spreadsheet authors often make efforts to improve the comprehensibility of spreadsheets, but barriers remain.

Common to both our findings and those of Ragavan et al. are the fact that despite the creators' intent and willingness to spend effort, there remain mismatches from the consumer perspective, some of which can be addressed through design interventions in the tools creators use. Thus, referring to the categorization and examples of gaps derived in our social media comment analysis, we created two designs that aimed to address many of the gaps we found, particularly those for which a technological, design-led intervention seemed appropriate. 

Our aim was to use these designs in conversation with content creators, to validate that such learner-facing and creator-facing interventions would be helpful and not introduce additional burdens, and additionally to elicit feedback about design implications for video tutorials and the creation process.

\input{tbl_designs.tex}

Our approach to using designs to facilitate the interview was based on the condensed form of design-led inquiry developed by Chalhoub and Sarkar \cite{chalhoub2022freedom}. In particular, the use of designs in the interview was not as a direct evaluation of specific features and their utility. Rather, we were interested whether adding tools for addressing potential gaps experienced by tutorial viewers at learning and creation time would be useful, acceptable, and effective. As in Chalhoub and Sarkar's scenario-based interviews, participants were shown these materials primarily help them understand future workflows, and invited to reflect on them.

Table \ref{tbl:designs} lists each gap targeted by our designs and summarizes which features might address them. We detail each design and its potential impact below. Not all gaps are addressed by these designs. \emph{Spreadsheet sharing} was not addressed, as sharing links to template workbooks mostly rectifies this gap. However, creators gave feedback on how Design 1 be improved to address this gap (Section \ref{sec:resultdprobe}). The \emph{Region and Language issues} gaps were not addressed, as they rarely occurred during Study 1. While other gaps left unaddressed occur more frequently (e.g., \emph{Unexpected behavior} and \emph{Adapting tutorials}), we focused on the most interesting subset of gaps that creators might have unique insights on. This kept the design probe portion of Study 2 tractable. 



\mysec{Design 1 - Interactive video tutorials within Excel}
\begin{figure}[ht]
  \includegraphics[width=0.7\textwidth]{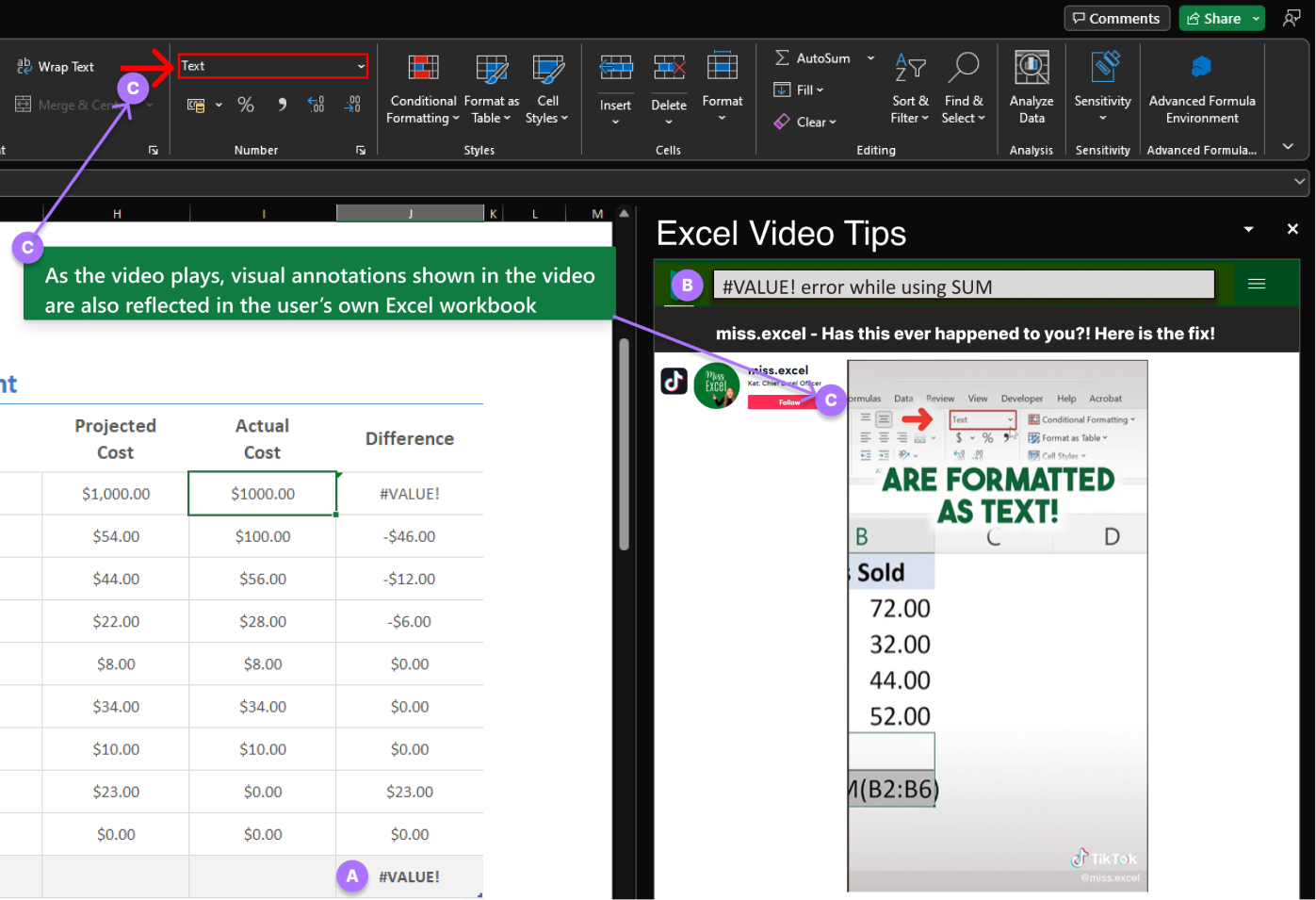}
  \caption{A frame of Design 1, which shows interactive video tutorials within Excel. \protect\ovcircle{A} Shows a user encountering a \#VALUE error. \protect\ovcircle{B} In response, the system queries the web for the issue and recommends relevant tutorials. \protect\ovcircle{C} As the tutorial plays, the user follows along with mirrored annotations within the application and completes each tutorial step (Figure \ref{fig:d1steps}).}
  \label{fig:d1annotated}
\end{figure}

\begin{figure}[ht]
  \includegraphics[width=0.7\textwidth]{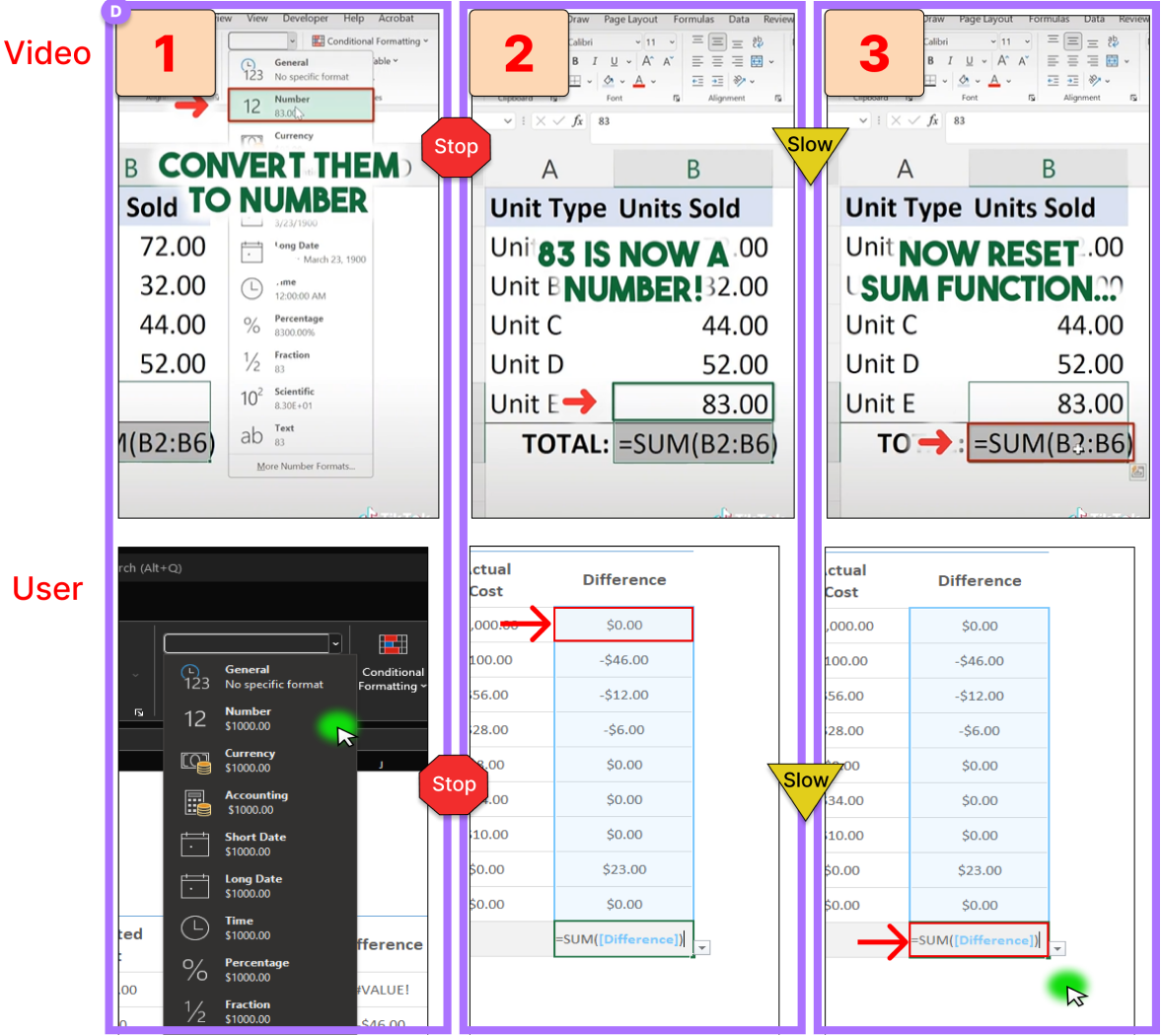}
  \caption{Example of the step-by-step nature of Design 1, where the video tutorial pauses and waits for the user to complete a step (from step 1 to step 2).}
  \label{fig:d1steps}
\end{figure}

The first design targeted several gaps for learners finding and applying online tutorials, by embedding tutorials into Excel and enabling connections between the tutorial and application (Figure \ref{fig:d1annotated}). When a user encounters an issue, such as an error, the system retrieves and displays relevant video tutorials in a sidebar add-in for Excel and helps users along their \emph{Learning journey} (Figure \ref{fig:d1annotated}, \protect\ovcircle{A}\protect\ovcircle{B}). The user selects a video to play within the sidebar. The user can walk through the steps shown in the video at their own pace; the video pauses at each step that required action (e.g., selecting from a drop-down menu, typing out a formula, etc.) (Figure \ref{fig:d1steps}, \protect\ovcircle{D}), which assists gaps that involve \emph{Missing steps} and \emph{Production issues}. Further, the visual annotations and animations shown within the video are reflected within the user's application (e.g., arrows pointing to the drop-down menu they need to select) (Figure \ref{fig:d1annotated}, \protect\ovcircle{C}). This design aims to address issues with the video going too fast for the learner to follow along, and bringing the tutorial within the context of the application they are learning about rather than having to context-switch between the video platform and their Excel application.

\newpage
\mysec{Design 2 - Identifying gaps for creators and providing alternatives} 
\begin{figure}[ht]
  \includegraphics[width=0.7\textwidth]{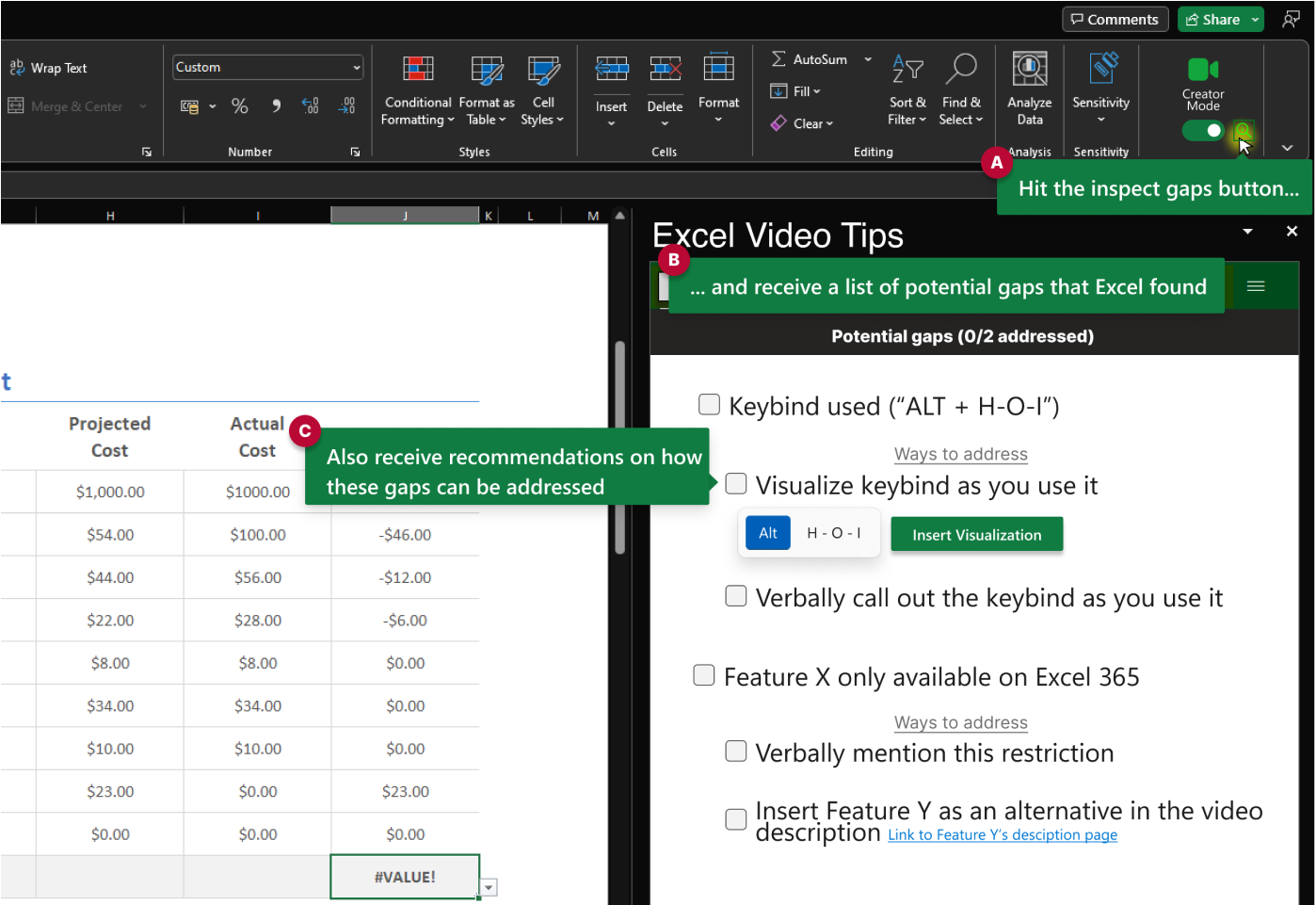}
  \caption{Design 2, which shows gaps detected in a content creator's video tutorial.  \protect\ccircle{A} As the creator works within a worksheet,  \protect\ccircle{B} a list of potential gaps is created that the user can view.  \protect\ccircle{C} Beyond the gaps detected, recommendations for addressing these gaps are also provided to improve video tutorial effectiveness. }
  \label{fig:d2annotated}
\end{figure}

The second design targeted the creation of tutorials themselves to assist content creators in identifying gaps in a tutorial, and providing options to bridge the gap (Figure \ref{fig:d2annotated}). A creator toggles a creator mode on when they are ready to record a tutorial (Figure \ref{fig:d2annotated}, \protect\ccircle{A}). The system monitors the actions taken by the creator and generates a checklist of potential gaps (Figure \ref{fig:d2annotated}, \protect\ccircle{B}). For example, if a feature or formula the creator uses is only available in a specific version of Excel (a \emph{Version issue} gap), it notifies them to inform their viewers, and generates a list of compatible versions for the creator to put in their video description (Figure \ref{fig:d2annotated}, \protect\ccircle{C}). If an alternate interaction is known to perform a similar task, a link to a documentation page could be inserted into the video description as well, possibly helping with \emph{Configuration question} and \emph{Unknown concepts} gaps. If a keyboard shortcut was used, it would visualize the keys pressed so that they could show viewers who might not be using sound, avoiding a potential \emph{Missing steps} gap. The creator can use this list to address potential gaps, to improve their tutorials.

\section{Study 2 methods: contextual interviews with content creators}
\label{sec:interviewmethods}

The second study in our investigation consisted of interviews with Excel video tutorial creators. Our questions covered the video tutorial creation process (e.g., picking a topic, editing, production choices), their frustrations with creating tutorial videos (e.g., the creation tools they use, frustrations with Excel, and with sharing their knowledge with learners), and the gaps their viewers face when watching and trying out the concepts found in the tutorials.

Participants were recruited from a creator engagement network program run by a global technology firm. Due to the personalized and long-term nature of this program, participants were not compensated specifically for participating in the study.


\input{tbl_creatordemos.tex}
We interviewed content creators (n=8, 5 women, 0 non-binary, 3 men) who publish video tutorials on the platforms we gathered comments from.
Table \ref{tbl:creatordemos} details participant information. All participants published on at least two of the platforms we investigated (Instagram, TikTok, and YouTube), with four publishing on all three. At the time of the study, participants had a mean of over 790,000 followers between the platforms (2 less than 100k, 3 between 100k and 999k, and 3 over 1 million followers).

We performed a 1-hour semi-structured interview over video call. Participants spoke about their experiences as a content creator for video tutorials. They also walked through an existing tutorial they had created to explain their creation motivation and decisions, and discussed how they could be supported better in this process. We grounded these interviews by asking participants to respond with specific examples drawn from their experience of creating a few of their self-selected video tutorials. However, our participants were not limited to these select videos and frequently referenced other video tutorials as examples.

The interviews covered the creative process for tutorial videos (e.g., \emph{``What is your creation process for your tutorials?''}), the motivation and decisions for video tutorials (e.g., \emph{``How do you select a topic for a specific tutorial?''}), frustrations with content creation workflows (e.g., \emph{``What struggles or frustrations do you have with the creation process that involve Excel?''}), viewer feedback, gaps, and how to address them (e.g., \emph{``How do you address viewer feedback?}), and tools for content creation and their limitations (e.g., \emph{``What features are missing that could help your tutorials be more effective teaching artifacts?''}).

For creative process, creators spoke about picking topics for the tutorial, creating spreadsheets, their motivation for being Excel tutorial creators, and how platforms affect video tutorials. When speaking about their frustrations, creators talked about their struggles with Excel, with the creative tools they use to record, edit, and publish their tutorials, with sharing knowledge, and the mistakes they have made. Finally, creators spoke about the gaps they see learners have with their tutorials (e.g., issues with different versions of Excel, following along with the tutorial, language of the tutorial and app, and finding the right tutorial) and how the creator may or may not address these gaps.  

As with the comment data, we used paired agreement to analyze and organize our creator interview responses into categories and subcategories. These are
detailed in Section \ref{sec:results}.


In a second phase of each interview, we introduced our two separate designs to participants to get feedback on whether features such as these would be useful in making video tutorials more effective for learners and ease any frustrations they had with the creation process. As mentioned previously, these probes were used as inspiration for the creators to identify any features or designs that might be useful in closing the gaps between learner and tutorial. Each design was presented through low-fidelity PowerPoint slideshow mockups, including annotated information on the motivation and goals for the design. Participants were also shown Figma \cite{figma} prototype images that demonstrated one user workflow. After each design was shown, participants were asked to reflect on what they had seen, including what they would find useful, how it would impact their creative process and the effectiveness of their tutorials, and what other features could help out in this area. The results are reported in Section \ref{sec:resultdprobe}.


\section{Study 2 results: Minding the gaps between the training and the platform}
\label{sec:results}

This section presents our interviews with Excel tutorial content creators. We explored how content creators address gaps we found from Section \ref{sec:resultgaps}, as well as the challenges they face. Further, creators spoke about their creation workflows for creating video tutorials and the frustrations they may have with creating video tutorials.
Finally, these creators reflected on designs aimed at mitigating gaps for creators and learners described in Section \ref{sec:dprobemethods}.

\subsection{How creators address comments about gaps}
\label{sec:creatorsaddressgaps}
Creators spoke about how they address comments about gaps, and the challenges they face in doing so.
For example, P2 said that when a question \emph{``is something that I can answer from the top of my head, the moment I see the comment, I try to answer it.''}
However, creators for whom content creation is not a full-time job have difficulty finding time to respond to questions and comments: \emph{``You know, like it's a simple, ruthless case of time and effort.''} (P1). 
P6 stated that the volume of comments they receive was manageable but subject to change, e.g., \emph{``subscriber numbers have sort of jumped up in the last six months and then getting more and more comments. So, it's getting a bit more onerous, just handling those''}, which might mean the creator might need to be more selective about the comments they respond to.

Some responses require careful thought or additional research and preparation to address. P7 noted difficulty with textual questions which were vague for the creator to understand the learner's exact issue the learner is having, and that sharing a short video clip of the situation might be better for clarifying the issue. P2 described a question that was interesting but would require a new video tutorial to address fully, but that viewers could be \emph{``very impatient, and they just expect that we create an answer right away, and I've had a case where the person wanted me to go on Zoom on a weekend to answer his question.''}


Videos on Instagram and TikTok are typically less than one minute in duration. Some gaps are inevitable with short form content, as \emph{``there's a tension between the short-form videos where you want to basically create something kind of cool and interesting in that maybe 9 or 15 seconds long. You don't necessarily want to start putting in loads of text at the beginning''} (P1). This might mean that responding to comments or publishing follow-up videos are the only solutions for short-form platforms.

When viewers spot mistakes in the tutorial, P3 said that they try to address it by thanking the viewer for letting them know and \emph{``if it's a major thing, we might change it, or even take the video down if it's like really something we'd screwed up.''}


Creators also used captions and video descriptions to pre-empt questions about keyboard shortcuts and version issues. In order to pre-empt viewer queries, P5 found it helpful to \emph{``write a really in-depth caption''} for the video, including a listing of all the functions used in the tutorial, which \emph{``has eliminated a lot of questions''} However, they acknowledge that this can have an undesirable consequence: comments increase engagement, and improve the video's performance on platform recommendation algorithms. P5 noted that questions drive engagement with the video which benefits the recommendation algorithms, so eliminating all questions may not be beneficial to a creator. In response, for example, P3 leverages the \emph{clarifying topics} gap by explaining their motivation for their specific approach to an issue, and then saying \emph{```Leave a comment below if you have a better way', to just keep the communication open.''}



\subsection{Creative process}
\label{sec:resultprocess}
Creators spoke about their creative process from idea inception to how publishing on their preferred platform affects their videos, and about their motivation for creating tutorials about Excel.

\mysec{Picking a topic} 
Creators had diverse sources of inspiration for choosing tutorial video topics.
P6 uses online forums like Reddit and the Excel Tech Community forum hosted by Microsoft to find common questions that Excel users are asking.
Several creators (P2, 3, 5, 7, 8) noted that they frequently read user comments on their published tutorials: \emph{``I think the first step is review the comments and the questions, find which are most commonly asked, and try to do videos about that''} (P7).
Some creators that also work in other professions (P1, 6, 8), like accounting or marketing, leverage their experience: \emph{``A lot of things I've shown is actually what I'm using in my real life at my work''} (P8). 
Some topic choices were motivated by the platform recommendation algorithms (P3, 4, 5, 8) by \emph{``looking at what are trending topics and searches''} (P3). P4 focused on finding popular functions, while others create tutorials for new Excel functions and features as they are released (P5, 6).

\mysec{Creating and sharing spreadsheets}
Creating spreadsheets to use in a video tutorial is a critical part of the process. These are the ``stage'' upon which creators teach, but are also requested by learners so that they can follow along.
Creators tend to generate \emph{``made-up data that I use to actually illustrate the examples''} (P1), often using random number generation in Excel.
Others use add-ins or external websites that generate fictitious data for them (P2, 3), since they \emph{``don’t want to deal with any type of copyright situations''} (P5) associated with using real data.
P8 uses their own interests to guide the process and add interest: \emph{``It's also showing my personality, right? I love Marvel, so I put Marvel there, so people are like `Oh that's so cool, I love Marvel too'''} (P8). As a civil engineer, P7 enjoys using construction data. 
Some reuse workbooks made for formal training courses (P3, 6). P3 notes that the tutorial determines \emph{``what fields [data columns] are necessary''} and based on the tutorial they \emph{``strip out everything that's not necessary, so it doesn't over clutter [...] it's kind of a balance.''}

Creators may also share the spreadsheets they created for their tutorial, which can address the \emph{spreadsheet sharing} gap. P7 said that they always provide a file \emph{``so that people can practice.''} However, issues with these uploads, such as partially completed spreadsheets, \emph{``can be a point of confusion''} for viewers trying to complete the tutorial step-by-step (P2). P5 noted that they withheld sample files until they \emph{``got more and more requests to add the files, even if they're like 6 cells of data that you can just type in yourself. You still get requests for them''}, which points to learners being less motivated to try out a tutorial they see in a video if learning materials to scaffold the experience are not provided.

\mysec{Editing and production}
Creators discussed rehearsing or scripting before recording their tutorials. 
Some practiced the features they were going to use during the tutorial to become \emph{``ready''} to record (P1). P3 limited their preparation to just an outline of \emph{``specific points I wanna cover and things I don't want to miss''}. P6 said they did not plan their tutorial lessons, preferring to \emph{``just record on the fly and then edit and upload from there.''}

After recording, they edit the videos to make sure the resulting tutorial is effective at teaching the viewer, but also getting their attention. P1 edits their videos by
to remove background noise and 
\emph{``zooming in on specific parts of the screen''} to focus the view on important elements in the UI. 
Creators may add a separately recorded voice-over and a video of the creator themselves, which can drive engagement with learners. P8 said that\emph{``people like [...] to see the actual person behind it''}. 

\mysec{Creation motivation}
Finally, creators talked about their motivation for creating tutorials for Excel. This ranged from sharing what they learned to building their own brand.
P2 sees their tutorials as \emph{``creating a repository for myself, and by putting them out there, hopefully helping someone else.''}
Some creators said that they had similar responsibilities in sharing their knowledge about Excel with their co-workers (echoing findings by Sarkar and Gordon \cite{sarkar2018spreadsheetlearning}), including hosting training sessions for their company. 
Creators that also have private training sessions or their own consulting work use the videos for branding and building an audience so that some viewers might choose to hire them for a deeper education in Excel.
Most creators stated that they enjoyed teaching others about Excel. P3 said they enjoyed \emph{``helping others and seeing other people save time and grow with Excel because it was a big thing in my career.''}
In a particularly poignant example, one Russian creator with friends in Ukraine relied on the creation process as an emotional outlet for the stress induced by the war: 
\emph{``I just couldn't stop thinking about it. It always was in my head. It was too much for me. My work was fun, but not fun enough to occupy my mind and I wanted to do something [...] just to distract myself, and also I guess help like any people too in my own way. [It was] something I can do? I mean, it's probably making no sense, but this is how I felt.''}

\subsection{Creator frustrations}
\label{sec:resultfrustration}
Creators spoke about frustrations they had with the platforms hosting the videos, the application covered by the tutorial (Excel), and with creation tools. Beyond these, creators had frustrations with engaging with learners, planning lesson content, and mistakes they make themselves.

\mysec{Frustrations with media platforms}
\begin{figure}
\includegraphics[width=\columnwidth]{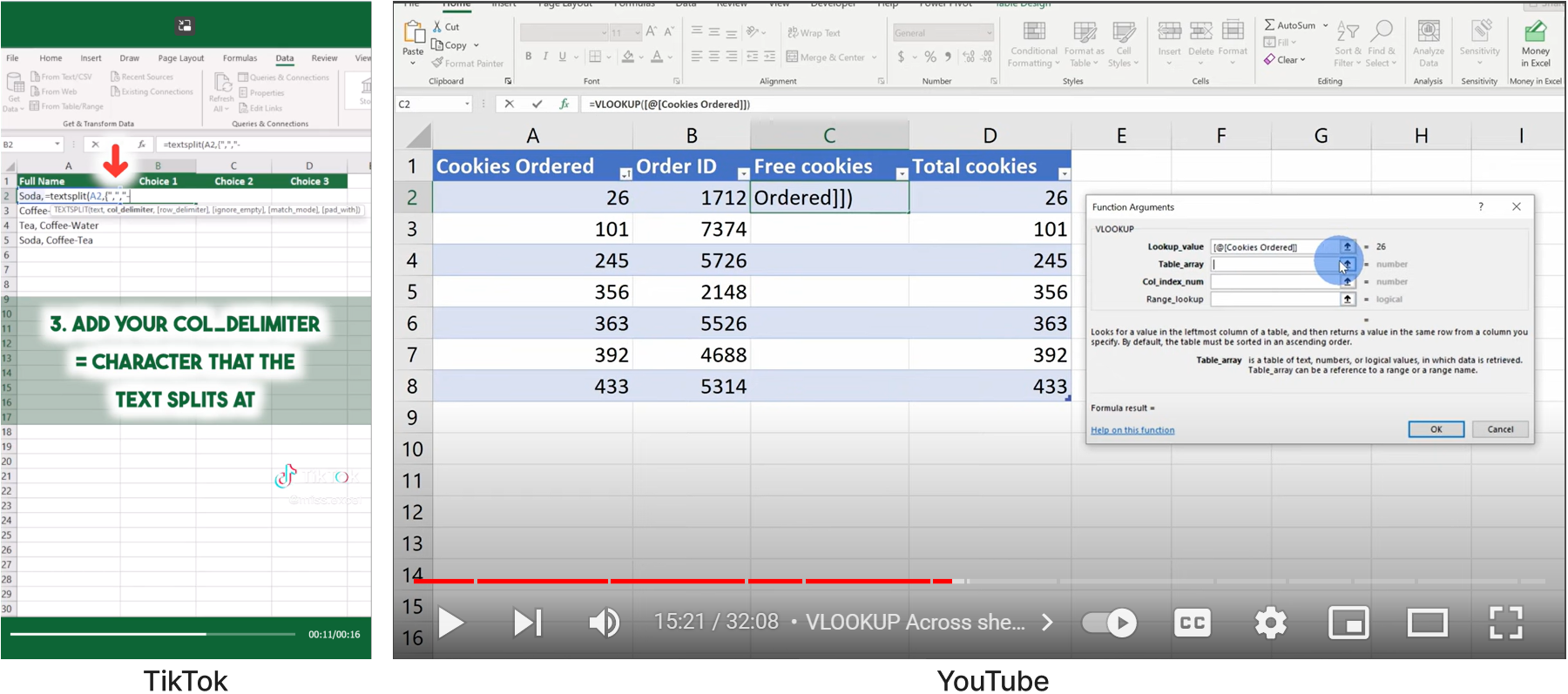}
\caption{A comparison of TikTok (left) and YouTube (right) viewport affordances for Excel tutorial videos.}
\label{fig:TTvsYT}
\end{figure}
Platforms strongly affect tutorial videos in many ways.
For example, TikTok, Instagram and the Shorts feature of YouTube require a 9:16 screen ratio for the video, which works well for mobile screens but can be a challenge for making videos about desktop applications as it does not allow creators to \emph{``show lots of different parts of the Excel screen''} (P1) (Figure \ref{fig:TTvsYT}). P2 noted that\emph{ ``producing Excel tutorials for Instagram or TikTok in vertical layout, depending on what we are teaching, is very challenging. We have to keep moving the [viewport] around a lot''}. P2 said that zooming is important since \emph{``some [Excel UI elements] are very small''}.
Re-editing a video to publish on multiple platforms was frustrating, especially for handling the different ratios between these platforms. 
P6 desired Excel to take over content tool duties by allowing for easy screen recording while visualizing where the barriers of their template of choice is: \emph{``if you could click a button, and it screen grabbed in Instagram or TikTok layout. So, you know where you've got to fit your screen into. That would be quite cool''} (P6).

The UI of the platforms can also interfere; for example, P4 noted that they only had about the middle 60\% of the screen to use on Instagram due to various UI elements like usernames and hashtags overlaid on the top and bottom.
Creators explained that the mostly white Excel grid clashed with the white buttons on TikTok. This obscures the like and share buttons from viewers, which lowers engagement. To fix this, P6 added \emph{``a black bar on the side just to make the white [buttons] stand out.''}


The length (duration) of the tutorial is also impacted. 
Short-form tutorials are commonly published on TikTok and Instagram, 
and require lessons to be simplified to fit within the time limit. Thus, \emph{``the tutorials not going to be as complex, and it's not going to go into as much detail''} (P3). P8 cited the recommendation algorithm as another consideration when cutting information: \emph{``the algorithm is not going to push a video if it's too long''}. So, they avoid putting alternative keyboard shortcuts in their videos.

This can lead to comments from viewers that the creator was going too fast for them to follow along. However, when P5 tried to slow down some of their videos, they saw that some viewers \emph{``won't even watch it if it's too slow''}, and thus they needed to find a \emph{``happy medium where most people get it''}. YouTube, in which longer videos are allowed and are more common, has an advantage for going into more depth. P8 stated that they \emph{``feel like YouTube is still the deeper and more educational platform. Especially for Excel.''} While longer tutorials give a creator the time to explain the details of concepts they are teaching, short-form platforms cause creators to think about whether their tutorial can be effective in less time. P5 critiques their own videos by asking \emph{``can someone learn it this quickly and is this the most efficient and optimized way, in a micro-learning sense, for them to learn it?''}.

Other differences involve the community norms and expectations around videos posted on a particular platform. TikTok and Instagram tutorials \emph{``have to be slightly clickbaity, and generally if you are not able to grab the attention in the first two or three seconds, then people will just scroll to something[else]''} (P4). P4 and P7 both noted that for these short-form platforms, it was beneficial to inject humor into the tutorial to help grab learner attention. 
It was recommended to add both voice-overs and captions to tutorials, giving viewers the option to consume the video with or without sound. \emph{``TikTok users usually have sound on, but other platform users might not''} (P8).
P4 believed that including voice-overs in their videos would make them more effective, but that \emph{``it's slightly more time-consuming''}. P4 worried that this effort would be wasted for their Instagram videos, where many viewers browse without audio.

Spreadsheet sharing is one way to engage with learners and close gaps. However, P1 noted that platform restrictions can get in the way, as they were unable to post links to these spreadsheets because comment links were restricted in TikTok/Instagram videos.

P6 puts detailed steps in the description of their videos so that viewers can follow the process shown in the video. However, they noted the need to remind viewers where to find this information, since larger descriptions are hidden by default on YouTube. As an alternative, they went on to imagine that the steps they took in the video would be reflected in the learner's Excel. This feature would allow learners to \emph{``actually see (the steps) being written as if you were typing it live within the Excel sheet, with a little video embedded''} (P6).

\mysec{Frustrations with Excel}
The first set of frustrations that creators had with Excel involved its interface (P1-3, 5, 6). 
Several creators noted difficulty with fitting the Excel UI within a mobile aspect ratio required by TikTok, Instagram, and YouTube Shorts (P1, 2, 6). P1 mentioned the challenge of \emph{``managing to fit  everything into the 9:16 aspect ratio''}.
One method for alleviating this issue is to use zooming within the application. However, Excel's native zoom functionality has inconsistencies and limitations: \emph{``we can zoom into the grid and the power query editor, but not the ribbon''} (P2). Zooming introduces the need to organize the Excel UI so \emph{``people can see what's going on''}, which may require editing the video footage, since zooming can introduce \emph{``all this dead space in between [the left and right parts of the UI]''} (P6).
Another UI issue involved the consistency of pop-up menus and dialog boxes. P6 noted these dialogs would sometimes pop up behind production elements (e.g., an inset video of the creator's face). This meant they had to move the element to an available part of the screen, add the transition for this movement, and move the element back after the dialog box had been dismissed.

Performance can also frustrate creators during the creation process. Slowness during computationally intense functions can be a \emph{``challenge''} for creators because \emph{``no one wants to sit there and watch Power Query load''} (P3). Editing footage of these slow times can prevent this issue, albeit placing a burden on the creator to do post-processing. Sometimes creators would edit out pauses caused by waiting for a function to run or even between lines that the creator would speak. It can be \emph{``pretty challenging to cut all of those (pauses) out''} which takes \emph{``manual effort''} from the creator to fix (P3).

Any crashes impact the creation process as once you reopen Excel it \emph{``can look a little bit different''} (P2) which impacts the visual consistency of the tutorial.



Versioning, a common learner gap, also frustrates creators: it causes \emph{``compatibility issues with different versions of Excel
''} (P4). Excel's language settings can cause issues for creators as they do for learners. P2 said they were having issues with making a VLOOKUP formula work for a tutorial. They noticed \emph{``this was right after a tutorial about changing the language (in Excel), and when I changed it to Portuguese, it changed the separators to decimals''} (P2), which caused issues with the formula.



\mysec{Frustrations with creation tools}
Video editing tools often involve a steep
\emph{``learning curve''} (P6).
Not all creators have video production expertise: \emph{``a lot of people creating videos aren't [technical]. So, for them to go learn video editing [...] is very overwhelming.''} (P3).



P1 found writing and positioning annotations in their videos to be tedious. This impacted the content they published, noting this as one of the reasons their videos \emph{``will sometimes lean towards having slightly fewer instructions,''} as creating keyboard shortcut annotations are \emph{``yet another thing to have to write, edit, and position''}. 



Highlighting, zooming, and cropping frustrated several creators (P1, 4, 8). 
P4 noted that it is \emph{``very difficult to even highlight something in the tools''} they use.
P1 wanted a feature that would allow the screen recording to follow their cursor around the screen \emph{``rather than having to squeeze everything into a small box''}, which might address this frustration.

One creator said that a lot of the work with creating video tutorials is making sure annotations and arrows are positioned correctly and that captions and labels are correct, but there was also great effort in \emph{`` keeping it (the tutorial) entertaining, so people will actually watch it''} (P5).

\mysec{Engaging with learners}
Creators also mentioned frustrations that involved the sharing of knowledge and video tutorials themselves (P1, 4-8). 
There is a balance of information that a creator can show in each tutorial video, since \emph{``getting all the necessary information into the video that the audience want and demand is a challenge''} (P1). P1 thought a source of this challenge was the differences between versions of Excel (e.g., Mac vs Windows).
P7 said that the time needed to produce several videos a week can be taxing, and that this is especially true for creators who also have full-time jobs on top of their content creation.

Even Excel videos are not immune to ``online trolls'': viewers who make deliberately antagonistic, counterproductive, insulting, or inflammatory statements in the comments. P4 said that they \emph{``can expect one at least one or two troll comments''} for a video, which might be discouraging for creators.


\mysec{Planning lesson content}
%
%
Trying to appeal to varying skill levels in the audience is also a challenge for creators, since learners can be at different stages of a learning journey. 
P4 said that appealing to all levels at once was \emph{``impossible''} and that there is a balance to be struck between brevity and including prerequisite knowledge for understanding a tutorial. For example, simply using a formal Excel table (a ``Ctrl-T table'' \cite{chalhoub2022freedom}) in a tutorial without first explaining how to do so, may cause viewer confusion. To avoid this confusion, P4 might \emph{``not use an Excel table in the example''}. However, this could cause issues when other viewers, who are aware of tables, would ask why the creator did not use one in the example (an instance of the \emph{Clarifying Choices gap}). This raises the question for creators of \emph{``where am I meeting people at? What is their current skill level and knowledge to get them to the next level?''} (P4). One creator addresses this issue by starting at the base knowledge needed to perform each step, and tells viewers to skip ahead in the video if they already have knowledge by saying \emph{``if you're not familiar with this function, start with me in this step''} (P7).

P8 mentioned that designing example real-world scenarios to demonstrate tutorial skills is difficult. However, one source of inspiration was their career using Excel: \emph{``a lot of times I get the ideas from things actually asked by my boss''} (P8).

\mysec{Mistakes they make}
Creators also spoke about the mistakes they made during production. P1 said that sometimes a video involves multiple takes due to simple mistakes. For example, when a creator clicks on the wrong tab when switching between Excel menus. This 
might not need a separate take for a long-form video on YouTube, but for short-form videos, a succinct workflow is necessary.

When a mistake happens in a recording, P2 noted a frustration with getting their Excel workbook into the state it was prior to the mistake: \emph{``otherwise people will see in the video that's not the same information [...] because I forgot something or misspelled something in my formula, and then I'm recording and need to go back and make it look as if there were no interruptions.''}

When creators see the mistakes they make, they can fix them with editing or by doing multiple takes. However, sometimes these mistakes still make it into the video. When they do, the viewer plays a role in spotting issues in the tutorials: \emph{``we're teaching the best analysts in the world, people that are trained to find errors in work, and so they're not only going to find errors in spreadsheets, they're going to find errors in content''} (P3). It can be difficult for creators to fix issues once they are published. P5 recommended automatic redirection from old videos, if supported by the platform, when creators update or correct issues in new videos.

\subsection{Design probe}
\label{sec:resultdprobe}
Creators were shown the design prototypes detailed in Section \ref{sec:dprobemethods}. We gathered feedback on how these designs might impact the effectiveness of video tutorials for learners. Further, we received feedback on how the designs might ease frustrations creators had with the creation process. Feedback is organized by design and by which gap it applies to.

\subsubsection{Design 1 - Interactive video tutorials within Excel} 
The first design allowed users to discover videos relevant to their current spreadsheet task (e.g., solving an error or using a feature). Users could then follow the steps seen in the video tutorial by interacting with their own Excel UI. During their interview and before seeing Design 1, P3 imagined \emph{``some level of interactive training''} would be beneficial. However, they felt it would be a challenge to create that type of experience \emph{``unless it's built into the application''} (P3). Design 1 was generally well-received by our participants, with the majority saying they would use such a design and P1 asking \emph{``when is it ready?''}.

In initial feedback, participants noted the design would be helpful for debugging errors and lower context switching between the video tutorial and the learner's app, mitigating the inefficient and cumbersome ``pause and play'' behavior observed by Pongnumkul et al. \cite{pongnumkul2011pause}. This seemed especially relevant for mobile consumption, since \emph{``watching it on a phone and then going to implement it involves a lot of back and forth''} (P4).

\mysec{Assisting the learning journey} 
The intervention aspect of the design was seen as useful by P5, since one of their motivations was helping learners when they needed it most. P5 said that help would be \emph{``coming at the right time, that's the biggest thing. Someone's panicking and my video could come in there and save the day. That's why I do what I do.''}


P7 remarked that this encouraged what they saw as the optimal, constructivist approach \cite{sarkar2016constructivist} to skill development: \emph{``the best way that someone can learn something is seeing it and doing it''}. They felt that without this practice, learners \emph{``forget what they saw in the video''} (P7).

P6 thought that the design would be most useful for beginner tutorials for learners just starting out by leveraging \emph{``the template files within Excel''}, but wondered if it would be as effective for more complex tutorials that they published.

However, P4 pointed out that relying on learners to search for topics for video tutorials could be difficult: \emph{``in most of the cases, the users don't really know what they want to do, how they can solve this problem. The problem is not that somebody doesn't know how to use XLOOKUP or SUMIF function. The problem is they don't know if there is such a function available''} (P4). The task-contextual nature of the system's recommendations were seen as helping to overcome these vocabulary problems; technical implementations of how this might be done have been explored by Fraser et al. \cite{fraser2019replay, fraser2020remap}.

P3 questioned whether in such a design users would actually be learning or just following the steps and annotations that pointed them where to go: \emph{``if learners are just following the arrows around then they're not really learning where that is, because they're just looking at the arrow instead of 'ohh it's on the Data tab', associating that memory versus just following the arrows''} (P3). Instead, P3 thought the design could be useful for Excel users who \emph{``need a refresher''} because they had not used certain features recently.

\mysec{Assisting production issues and missing steps}
P7 described a common workflow for learners watching short-form tutorials: watching the video several times to understand all the steps. P7 saw a benefit of this design in helping users manage the pace of a tutorial in a more fine-grained manner, which would give each learner a personalized time frame within which to understand and process the tutorial.
P5 stated that \emph{``people get overwhelmed by all the buttons''} within Excel, and that this could help prevent that by focusing the learner's attention.
P8 believed that this design could be useful for finding these UI elements that were often \emph{``hidden somewhere''} which can take learners time to find.


One challenge involves the screen real-estate that a video would get. P3 thought that playing a mobile-form video in a sidebar would likely be fine, but a full-screen YouTube video might be too small for a learner to see.

\mysec{Assisting spreadsheet sharing} 
P2 mentioned another potential
challenge involving the data discrepancies between the learner's workbook and the expected format needed for the tutorial.
This might make it difficult to apply a tutorial to the learner's situation. P8 offered up a solution for this issue: \emph{``you can give them a free template, and then it's the exact thing I do in my tutorial, so they can just watch it and do it.''} However, this does not solve the \emph{Adapting tutorials} gap, as the learner must still transfer this new knowledge to their own context afterward.

\mysec{Assisting version issues}
P3 noted that the \emph{Version gap} involving different versions of Excel would still be present in this design. One solution is to only present results that are relevant to the learner's current version of Excel. 

\mysec{Assisting language issues}
Language gaps could exacerbate the difficulty of finding relevant videos via text search \emph{``because there are a lot of users for whom the first language is not English, but they are using Excel in English...
in that case, Excel users are not able to find the right function''} by searching (P4). 
Instead, P5 thought you could replace search by having the top issues learners face have \emph{``pre-linked up''} video tutorials that related to the functions and features being used by the learner, and that were vetted as being relevant to the language localization of the user's Excel installation.

\subsubsection{Design 2 - Identifying gaps for creators and providing interventions}
The second design targeted the creation of tutorial videos for Excel and helping creators close gaps. The system presented in Design 2 monitors the steps creators took within Excel during the recording of a tutorial and warn them when there might be potential gaps (e.g., keyboard shortcuts are used or the function they use is only for a certain version of Excel). The system collates recommendations for closing those gaps. This design was also generally well-received by the creators we interviewed, particularly with helping out the \emph{Versioning} gaps, as \emph{``exactly which functions and which versions are something that I'm not going to be able to memorize any time soon''} (P1).

\mysec{Assisting production issues and missing steps}
While some creators thought adding visualizations within Excel would be useful (e.g., to show the keys pressed by the content creator), P4 noted that alternatives to this feature already exist, like accessibility features built into operating systems and tools. However, they felt that this design might still be useful as a reminder to enable them. They also expressed interest for Excel to handle all the recording functionality, similar to Microsoft PowerPoint's inbuilt screen recording feature, as it removes the need for \emph{``any other software''} (P4). For alternative keyboard shortcuts (e.g., for Mac versions of Excel), P3 thought that the automatic annotations for these offered by the tool would be \emph{``super helpful, because otherwise the creator or the video editor has to go look those up''} if they were unaware of alternatives. The ability to modify the location and appearance of these keyboard shortcut visualizations was seen as key for creators, as \emph{``if we end up like zooming on another part [of the screen], it [can be] kind of cropped out, and then we would have to edit it in anyway''} (P5).

P8 thought this design could also generate a textual description of steps the creator took during their tutorial. This was so learners could use these steps to follow along: \emph{``it would be also awesome if you recorded, and then it creates the steps, so we can just copy it and paste it [into the description]''} (P8), and thought AI might be able to help with this.

\mysec{Assisting version issues}
P6 believed that detecting which versions of Excel work for a certain feature was \emph{``bit of a minefield''}. Even within the same version of Excel, there are several different channels (e.g., different subscription tiers, or beta and general release channels), some of which might have features disabled. Users need a way to easily find their current version, but also what channel they are on. 
P6 noted that \emph{``lots of creators are doing videos using the insider [beta] channel, which isn't available for every user''}, so a feature to remind creators when they are on these channels of Excel and notify their viewers could be useful.

\mysec{Assisting learning journeys}
P7 said the description generation feature could help the discoverability of their tutorials: \emph{``you can have a great video, but if you can't make a description that the people can find your video, you can't help anybody like that''} (P7). P2 had a related suggestion to extend this feature to include a list of all features used within a specific tutorial for \emph{``SEO (Search Engine optimization) purposes''} so learners could find relevant tutorials. 

\mysec{Assisting language issues}
P7 wanted a \emph{``translation of the Excel functions to my language''} as a feature, which may be helpful to the issue that P4 pointed out in Design 1 where language gaps may cause issues with the ability to search and find related videos when you might not know the English word needed to find it. P2 also found it potentially useful as a non-native English speaker making tutorials in English: \emph{``English is not my first language, so I'm constantly translating in my head... if at the end you get this checklist to go through and find a couple of things that you can improve or insert, I find that very insightful, and I can see this being useful.''} (P2).

\mysec{Closing gaps makes effective tutorials}
Closing these gaps is useful for creators. When learners try out a tutorial and it \emph{``doesn't work for them, then they might not even look at your other videos''} (P4). This can weigh on creators who want their viewers to find their tutorials useful and do not want them to be blocked by gaps. 
P1 said they \emph{``feel a bit bad when I put a video out there''} about a feature that a viewer might be excited to try out, and it has \emph{``not worked for them''} (P1).
The features in this design \emph{``would close those gaps and lead to less frustration''} (P1). P5 said if Design 2 was implemented it would be helpful for learners and creators: \emph{``it would totally step up the game, really from my own perspective as a teacher, having that all at my fingertips just makes my life so much easier''} (P5).

\subsubsection{Design probe summary}
Table \ref{tbl:designsResults} modifies Table \ref{tbl:designs} to show the extent that each design may address each gap, grounded in feedback received from creators. 
\input{tbl_designsResults}
The two designs were found to be useful for closing some of the gaps we discovered in Study 1 to increase video tutorial effectiveness. Design 1 is a learner-side intervention that assists learners on their learning journey, intervening with tutorials when learners most need it. Design 1 encourages learners to try what they see, which can help maintain knowledge gained from the video tutorial. Design 2 is a creator-side intervention that helps keep tutorial creators aware of the potential gaps their tutorial might have and recommends potential solutions to close them. This reduces some of the manual work that creators already do to make their video tutorials accessible to learners.

Creators also provided insights on how these two designs might address gaps we originally did not consider. For example, Design 1 would be improved by generating and sharing template spreadsheets to assist in \emph{Spreadsheet sharing}. Design 2's description generation can assist in the \emph{Learning journey} gap by producing more informative descriptions, which helps learners find the right video.
However, these designs did not completely close gaps. While Design 1 sought to address the \emph{Learning journey} gap, creators questioned if learners would use it as a learning tool or simply copy the steps without gaining much knowledge. Creators detailed that Design 2 needed to be enhanced to completely address \emph{Production and Version issues} gaps.






\section{Discussion}
\label{sec:disc}

\subsection{Comparison with related work}
Our findings further substantiate many results documented in prior literature. For example, we found that following along step-wise with the tutorial was both a common user behavior and a source of gaps. As found in previous work \cite{guo2014video,kiani2019beyond,chi2012mixt,pongnumkul2011pause}, users were unable to follow along if the video had production issues, the video was too fast, missed steps, or assumed too much prior knowledge. We further identified that differences in software versions was an extremely common source of issues. This is a particular problem for software such as Excel, which not only has different versions for different operating systems, but also multiple supported versions on the same platform as per the vendor's software licensing business model. This suggests that video tutorial websites could incorporate filters for software versions, or the user's software version information could be used to automatically enhance the tutorial search query (as previously explored by Fraser et al. \cite{fraser2019replay,fraser2020remap}).

We found that commenters wanted to clarify choices made by the creator (Section~\ref{sec:resultgaps}). This corroborates some of the findings of Ragavan et al.'s study of spreadsheet comprehension \cite{srinivasa2021spreadsheet}, who also found that choices of specific functions, unexplained parameters or ``magic numbers'' were sources of comprehension barriers. We found that recall was a challenge (Section~\ref{sec:learner-driven-gaps}), similar to problems addressed by Grossman et al. \cite{grossman2007strategies}.

\subsection{What we owe each other}
A striking phenomenon emerging from both user comments and creator interviews was the nebulous nature of the social contract between tutorial creator and consumer. The tutorial exchange is mutually beneficial; creators typically benefit from increased visibility, status, authority, and ultimately sales of formal training offerings, advertising revenue, etc. In turn, consumers benefit from a wide variety of tutorials at different levels, available to them at no marginal cost, and access (albeit limited) to the expertise of the creators. Creators often felt a tension between responding to requests made by commenters and managing their own time and agenda. They often felt a sense of obligation, duty, or etiquette, to engage with commenters and satisfy their demands. Due to the sheer scale of platforms such as YouTube, they often simply cannot respond to every request. Moreover, not every request is intelligible, and in some cases, the request is so trivial that creators describe it as ``lazy'', the implication being that the commenter ought to have been able to solve the issue themself. On the other hand, they appreciate any engagement because it helps to build a sense of community, and raise the profile of the videos on the platform's algorithmic recommendations.
Creators described bringing on additional staff to their team to help with addressing user requests. A platform such as Discourse \cite{KindelDiscourse2017}, which assists in the analysis of MOOC forum discussions, can be useful for creators to analyze comment data on their videos to better diagnose what gaps their viewers are experiencing.

User comments indicated a tension between acknowledging the tutorial ecosystem as a free resource, versus the potential for (or entitlement to) deeper, personalized engagement with the creators. User comments ranged from pure comments of support, to suggestions and feedback on the tutorial itself, to a variety of requests: for help with specific questions, with their own spreadsheets, to recommend alternative tutorials, and to create entirely new videos (sometimes in an entirely different language!). Previous research recommended creating content in multiple languages to make instruction more inclusive, and that simple translation might not be enough \cite{RUIPEREZVALIENTE2022104426}. We found evidence of this, since the names of Excel functions in different languages may not be direct translations of each other. These requests show a widely divergent multiplicity of perspectives regarding what one should ``pay in'' to the free tutorial ecosystem, versus what one should receive in return. Mismatches in expectations may cause confusion: for instance, a user may post a trivial question, not because they wish to have it answered, but out of a genuine desire to demonstrate engagement with the community and boost the profile of the video in the algorithmic rankings (the equivalent of small talk), but this may appear to the creator and other viewers as ``lazy''.

\subsection{Design implications}
\label{sec:discfish}
The gaps and themes derived from the comments and creator interviews indicate several design implications for making tutorial videos more effective. We also note the recent progress in Large Language Models (LLMs) as a way to enable future designs for better feature-rich software learning and tutorial creation, or even help eliminate some gaps. Models such as OpenAI's GPT-4 \cite{openai2023gpt4} and Meta's LLAMA 2 \cite{touvron2023llama} may contribute to how creators design tutorials and how learners use them.

\mysec{Helping learners find relevant tutorials} Many comments indicated that video tutorials are important to a person's learning journey. There is a need to think about video tutorials as more than individual artifacts that stand in isolation, and rather to consider the breadth of resources that can be leveraged by learners throughout this journey and how each of them might build on and interplay with the others. Many commenters requested alternate tutorials, playlists of videos that they could follow from beginner to advanced levels, and asked creators to help them adapt a tutorial to their own context. 

Therefore, tools aimed at tutorial search and recommendation should consider not only the current context of the user (e.g., what task are they doing now and what tutorial would best help them do it?), but the journey the learner is on. This includes prior tutorials used, previous knowledge gained, and intended future learning trajectory (e.g., beginner tutorials explain more prerequisite knowledge needed to understand the tutorial). Several creators raised the issue that viewers may be at different levels of knowledge. Creators need to balance the tutorial to give enough information that most viewers can effectively learn. 

Further, many commenters request help for their specific situation, which might be addressed by presenting them with videos that are closest to what they need, and meets them at their current level. Here, for example, LLMs might assist in adapting existing tutorials to the unique context of the learner. If a learner wants to apply a generic tutorial to their own spreadsheet, an LLM could, for instance, be given as input the transcript and description from the video, and asked to derive a tailored set of steps for the learner to apply in their own spreadsheet with their own data and Excel installation idiosyncrasies taken into account.


\mysec{Helping creators handle platform restrictions} One major frustration for creators who publish videos for mobile-first platforms, like on TikTok and Instagram, was fitting Excel's dense desktop UI into the small vertical screens of mobile devices. Creators have to zoom in and pan around the interface for viewers using a mobile phone. This movement can confuse a learner who is both trying to follow along and retain what they are learning. Applications could support mobile-first video tutorials by offering alternative interface layouts, which better accommodate mobile viewing without excessive zooming or panning.

\mysec{In-context, interactive tutorial videos} Learners had many difficulties following along with tutorials. The video might have been perceived as progressing too quickly, which can be exacerbated by short-form videos on TikTok and Instagram. Another gap is that some steps were difficult to understand or even missing. Other comments alluded to difficulties with finding the ``right'' tutorial when needed.

This highlights the potential for an integrated experience within an application for discovering and following video tutorials, such as Design 1 in Section \ref{sec:dprobemethods}. This design combines many previously considered aspects of software help, such as proactive help \cite{xiao2004empirical}, context-sensitive querying \cite{fraser2019replay}, task-centric and stencilling interfaces \cite{lafreniere2014task,kelleher2005stencils}, and automatic synchronization of video to interface actions \cite{pongnumkul2011pause}.



Besides the numerous technical challenges for this design idea to be realized, there is also the question of whether such a design can be effective in helping learners gain knowledge versus \emph{``just following the arrows''} (P3).
This again raises the balance of holding the learner's hand through the steps, versus letting them explore the UI to gain practice at navigating it. However, for some learners, such hand-holding can alleviate many of the gaps experienced when trying to learn from tutorials for feature-rich applications.


\mysec{Detecting gaps and generating solutions for creators}
Another set of opportunities lies in targeting the creation of tutorials themselves, assisting content creators in identifying gaps in their tutorials and provide options to bridge them, as illustrated in Design 2 in Section \ref{sec:dprobemethods}.

For instance, this could help close the version issues gap, since remembering the limitations of each feature or version a challenge for creators, which they will \emph{``not be able to memorize any time soon''} (P1), 
and requires the creator to repeatedly research this information (P3). 
Creators noted some of the features were also helpful for discoverability of tutorials and engagement with learners. P7 noted that description generation features could help the discoverability of their tutorials for learners, while P8 suggested that a recording tool could generate the steps the creator took during their tutorial in text, which could be pasted into the description, so learners could use them, and platforms could better index the video. LLMs could help creators write detailed video descriptions which include details like version requirements for formulas and features, which would alleviate this burden on content creators and help platforms more effectively index their video tutorials.

\mysec{How LLMs can enable designs for more effective learning}
Current progress in LLMs might assist these realizing the design goals we showed to creators, and may even help eliminate some gaps altogether. 

Beyond the design implications we discuss above, LLMs might assist in other gaps we found as part of our social media comment analysis and creator interviews. For spreadsheet creation and sharing, LLMs can generate relevant spreadsheet data for both creators to use in their tutorials and for learners to practice the tutorial they see.

Unknown concepts can quickly be explained by LLMs, rather than depending on creators responding to comments on tutorial videos uploaded to social media. 
To help learners practice the tutorials to maintain knowledge gained from watching tutorials, LLMs might be able to generate alternative tasks that relate to the topic of a video and expand the amount of learning resources available to learners. More broadly, LLMs could generate a wide range of learning resources to supplement tutorials, like explanations of concepts, practice questions, and customized practice data sets. They could also translate materials to other languages to expand access.

These are a few potential applications of LLMs that relate to the design implications we have discussed here. 
However, further research is needed to discover the potential applicability of LLMs to feature-rich software tutorials and learning which includes AI-generation of tutorial steps and code, or even the videos themselves.

Moreover, LLMs are not a panacea: they suffer from various technical and ethical issues such as hallucinations and biases. While large language models offer promising capabilities for improving video tutorials, there are also important limitations and risks to consider. One major concern is potential biases that could be amplified through LLMs. As models are trained on imperfect human-created data, they may inadvertently propagate harmful stereotypes or assumptions. This could lead to unequal treatment or exclusion when generating personalized instructions or practice exercises. Rigorous testing and mitigation of biases would be essential.

Another key risk is that of hallucination, where the model confidently generates plausible but false or nonsensical outputs. This could clearly be dangerous in an educational context, where accuracy and appropriateness are paramount. Learners could be misled by incorrect steps or explanations. Ongoing advances in grounding LLMs in external knowledge and reasoning can help, but must still be validated. Overall, while LLMs offer useful potential for video tutorials, researchers must carefully consider countermeasures for bias, hallucination, and other issues to ensure educational effectiveness and ethical application.


\label{sec:discdesign}

\subsection{Study scope and limitations}
\label{sec:limitations}

Our data collection and interviews focused only on Excel video tutorials and Excel content creators. Other domains may have different needs and challenges (for example, multiple-camera setups for artists \cite{drosos2022}). The gaps we found may also exist in instructional livestreams \cite{drosos2021}, but our data does not directly address these. Creators we interviewed may be subject to hindsight bias as they reflected on previous videos and their initial creative processes. We nonetheless believe many of our results could apply to video tutorials for other feature-rich software such as PhotoShop, AutoCAD, etc. 


Our video selection process may be affected by selection bias and by the platforms' own search algorithms. Analyses of online communities are subject to self-selection bias; participation in comments on online platforms such as YouTube, TikTok, and Instagram is skewed towards spreadsheet users with an intrinsic interest in learning and in developing their technical skills. This may limit the representativeness of our sample.

We modeled our search to emulate what an Excel learner would see in their own search results. In our analysis we found that for some learners that language is an important aspect that may result in gaps, but we only collected English comments, because of the linguistic abilities of our research team. A few of our creators discussed language as a potential issue for video tutorials, but further investigation in this area is needed. We looked at three popular platforms for publishing video tutorials, but LinkedIn and learning specific platforms (e.g., SkillShare, MOOCs) may have other challenges not covered by our analysis. We analyzed comments posted on relevant tutorials, but did not interview commenters to confirm their actual issues. Hence, there is some ambiguity in interpreting the gaps that appear in each comment, and while we made an effort to negotiate agreement between multiple researchers to reduce subjectivity, we cannot be certain of the root cause issue in every comment.

The interview component of our study has limitations common to much of qualitative research. The quality of the insights elicited depends on the researchers' individual skills and might be influenced by their personal biases. Inexperienced interviewers may fall short in posing timely inquiries or delving beyond surface narratives, thus failing to elicit salient empirical evidence from respondents' accounts \cite{koskei_role_2015}. The strength of interview data is dependent on the interviewer's skill \cite{kajornboon_using_2005} and the nature of the questions they ask \cite{birmingham_using_2003}. In order to mitigate potential biases stemming from variability among interviewers, a single researcher possessing expertise in conducting non-leading interviews from a neutral standpoint carried out all interviews. This researcher was trained in consistently applying open-ended, unbiased questioning techniques so as to avoid inadvertently influencing subjects' responses.

Interview studies are commonly subject to self-reporting bias~\cite{Jupp2006}. Certain respondents may have provided responses that were not entirely accurate due to limitations in their recall of specific details. Conversely, a subset of participants might have exhibited response alterations stemming from apprehensions regarding the interviewer's evaluative stance, thereby aligning their answers with a desired self-presentation. This phenomenon is exemplified by the potential impact of sociocultural variables, such as ethnicity, on the propensity of distinct societal cohorts to furnish certain types of responses \cite{c_n_trueman_structured_2015}. To maximize validity and minimize self-reporting bias, we avoided leading questions and relied on open-ended questions, inviting participants to provide in-depth answers in their own words. Some of our participant answers were less detailed than others. Still, we prompted participants to give full answers to all questions. We ensured that participants' responses were grounded with respect to specific events and concrete artifacts (e.g., their own video tutorials), to improve recall.


\section{Conclusion}
\label{sec:conclusion}
We present an analysis of gaps encountered by learners when learning from a video tutorial. Many of the gaps represented a need for the feature-rich application to adapt to the content platform that shows the tutorials. For instance, if people are learning on mobile video platforms like TikTok, it would benefit an application's users if it supported the creation of tutorial content for that format. Moreover, the learning journey gap represents a need to look beyond video tutorials as independent artifacts, and instead to consider the range of resources available to learners. Further research is needed on how to guide learners to find and adapt these video tutorials to their own context to best support their learning journey and leverage the quickly growing areas of LLMs. We hope that this research inspires new designs that help creators ``mind the gap'' so that learners can benefit more from their tutorials.

\bibliographystyle{elsarticle-num} 
\bibliography{references}

\appendix

\section{Tables of videos we analyzed}
Below are the videos we used to gather comments for our analysis on YouTube (Table \ref{tbl:videos}), TikTok (Table \ref{tbl:ttvideos}), and Instagram (Table \ref{tbl:igvideos}).
\newpage
\input{tbl_videos.tex}

\input{tbl_ttvideos.tex}

\input{tbl_instavideos.tex}





\end{document}

%% file: tbl_findings.tex
\begin{table*}

\caption{Themes and their supporting evidence. Each row denotes a theme found during our investigation, with a column for each individual study component: the comment analysis, creator interviews, and design probe. A dot indicates that we obtained evidence towards our understanding of a theme. For example, the theme ``spreadsheet sharing'' has dots across all three columns, indicating that this theme was encountered in all three study components. The theme ``configuration question'' has only one dot under ``comment analysis'', indicating that this theme was encountered during the comment analysis but not during the creator interviews or design probe.}
\label{tbl:findings}

\begin{center}
\begin{tabular}{{@{}lccc@{}}}
Theme
& \shortstack{Comment\\analysis} 
& \shortstack{Creator\\interviews} 
& \shortstack{Design\\probe}
\\
\midrule
\emph{Video tutorial gaps} & & & \\
\multicolumn{1}{r}{Spreadsheet sharing}
& \hfil\yellowdot 
& \hfil\orangedot 
&  \hfil\orangedot
 \\
\multicolumn{1}{r}{Production issues} 
& \hfil\yellowdot 
& \hfil\orangedot 
& \hfil\orangedot 
\\
\multicolumn{1}{r}{Missing steps} 
& \hfil\yellowdot 
& \hfil\orangedot 
& \hfil\orangedot 
\\
\multicolumn{1}{r}{Clarifying choices} 
& \hfil\yellowdot
& \hfil\orangedot 
&  \\
\multicolumn{1}{r}{Learning journey} 
& \hfil\yellowdot 
& \hfil\orangedot
& \hfil\orangedot\\
\multicolumn{1}{r}{Adapting tutorials} 
& \hfil\yellowdot 
& \hfil\orangedot 
&  \\
\multicolumn{1}{r}{Maintaining knowledge} 
& \hfil\yellowdot 
& \hfil\orangedot 
& \hfil\orangedot \\
\multicolumn{1}{r}{Unknown concepts} 
& \hfil\yellowdot 
& \hfil\orangedot 
&  \\
\multicolumn{1}{r}{Region issues} 
& \hfil\yellowdot 
& \hfil\orangedot 
&  \\
\multicolumn{1}{r}{Language issues} 
& \hfil\yellowdot
& \hfil\orangedot 
&  \hfil\orangedot \\
\multicolumn{1}{r}{Version issues} 
& \hfil\yellowdot 
& \hfil\orangedot 
& \hfil\orangedot \\
\multicolumn{1}{r}{Configuration question} 
& \hfil\yellowdot 
&  
&  \\
\multicolumn{1}{r}{Unexpected behavior} 
& \hfil\yellowdot 
& \hfil\orangedot 
&  \\
\hline
\emph{Creation process} & & & \\
\multicolumn{1}{r}{Picking a topic} 
& 
& \hfil\orangedot 
&  \\
\multicolumn{1}{r}{Creating spreadsheets} 
& 
& \hfil\orangedot 
& \hfil\orangedot \\
\multicolumn{1}{r}{Motivation} 
&  
& \hfil\orangedot 
&  \\
\hline
\emph{Creator frustrations} & & & \\
\multicolumn{1}{r}{Media platforms} 
& 
& \hfil\orangedot 
& \hfil\orangedot \\
\multicolumn{1}{r}{With tools} 
&
& \hfil\orangedot 
& \hfil\orangedot \\
\multicolumn{1}{r}{Engagement} 
&  
& \hfil\orangedot 
& \hfil\orangedot\\
\multicolumn{1}{r}{Making effective videos} 
& 
& \hfil\orangedot 
& \hfil\orangedot  \\
\hline
\emph{Design implications} 
& \hfil\yellowdot 
& \hfil\orangedot 
& \hfil\orangedot \\
\bottomrule
\end{tabular}
\end{center}
\end{table*}

%% file: tbl_commentgaps.tex
\begin{table*}

\caption{Gaps identified through thematic analysis of corpus of 360 online comments (comments can exhibit multiple gaps)}
\label{tbl:gaps}
\begin{scriptsize}
\begin{center}
\begin{tabular}{@{}lp{95mm}c@{}}
Gap code & 
Description & 
\shortstack{Comment \\count} \\
\midrule

Spreadsheet sharing 
& Requests a sample tutorial spreadsheet or file that the creator used 
& 14\\

Production issues 
& Issues centered around the video production of the tutorial 
& 35 \\

Missing steps 
& Video tutorial skips over steps relevant to completing the tutorial 
& 26 \\

Clarifying choices 
& Why the creator chose a certain strategy or function 
& 37 \\

Learning journey 
& Finding new, alternate, or specific tutorials to further learning 
& 104 \\

Adapting tutorials 
& Viewer has specific needs in their spreadsheet that need more information to use the tutorial 
& 78\\

Maintaining knowledge 
& Expressing difficulties in remembering the lessons learned in the tutorial 
& 14 \\

Unknown concepts 
& Certain concepts or details needed to understand elements of the tutorial are unknown 
& 70 \\

Region issues 
& Region causes difficulties with following along with the tutorial 
& 5 \\

Language issues 
& Requests information or tutorials that relate to a language 
& 4 \\

Version issues 
& Excel version causes an issue with following along
& 61 \\

Configuration question 
& Impact of application configurations 
& 4 \\

Unexpected behavior 
& Learner is receiving behavior counter to what the tutorial shows 
& 65 \\
\bottomrule
\end{tabular}
\end{center}
\end{scriptsize}
\end{table*}

%% file: tbl_designs.tex
\begin{table*}

\caption{How Design 1 and Design 2 might address gaps we found in the comment analysis. Under each design there is the feature description that would address the gap. Each row denotes a gap found during our investigation, with a column for each design. 
}
\label{tbl:designs}
\begin{footnotesize}
\begin{center}
\begin{tabular}{{@{}cp{45mm}p{45mm}@{}}}
\shortstack{Gap code} 
& \shortstack{Design 1 \\ Interactive video tutorials} 
& \shortstack{Design 2 \\ Identifying gaps, alternatives}
\\
\midrule
\multicolumn{1}{l}{Production issues} 
& Stop-and-go video tutorial steps (Figure \ref{fig:d1steps} \protect\ovcircle{D})
& Annotation recommendation (Figure~\ref{fig:d2annotated} \protect\ccircle{C})
\\
\multicolumn{1}{l}{Missing steps} 
& Stop-and-go video tutorial steps (Figure~\ref{fig:d1steps} \protect\ovcircle{D})
& Annotation recommendation (Figure~\ref{fig:d2annotated} \protect\ccircle{C})
\\
\multicolumn{1}{l}{Learning journey} 
& Automatic retrieval of relevant video tutorials based on user actions (Figure \ref{fig:d1annotated} \protect\ovcircle{A} \protect\ovcircle{B})
& N/A
\\
\multicolumn{1}{l}{Maintaining knowledge} 
& Enabling learners to practice what they see, step-by-step (Figure~\ref{fig:d1steps} \protect\ovcircle{D})
& N/A
\\
\multicolumn{1}{l}{Unknown concepts} 
& Learner-driven search to find relevant video tutorials (Figure~\ref{fig:d1annotated} \protect\ovcircle{B})
& N/A
\\
\multicolumn{1}{l}{Version issues} 
& Recommending videos relevant to the learner's current version (Figure~\ref{fig:d1annotated} \protect\ovcircle{C})
& Detection of features restricted to certain versions and recommending descriptions (Figure~\ref{fig:d2annotated} \protect\ccircle{C})
\\
\multicolumn{1}{l}{Configuration question} 
& N/A
& Generating information for alternative interactions (Figure~\ref{fig:d2annotated} \protect\ccircle{C})
\\
\\
\bottomrule
\end{tabular}
\end{center}
\end{footnotesize}
\end{table*}

%% file: tbl_creatordemos.tex
\begin{table*}
\caption{Creator self-reported demographics}
\label{tbl:creatordemos}
\begin{scriptsize}
\begin{center}
\begin{tabular}{@{}llllllll@{}}
ID & YouTube & TikTok & Instagram & Videos & \shortstack{Videos\\per month} & Audience & Other profession \\
\midrule
P1 & Yes & Yes & Yes & 300 & 10 & Novice, Intermediate, Expert & Digital marketing \\
P2 & Yes & No & Yes & 100 & 2 & Intermediate, Expert & Excel consultant \\
P3 & Yes & Yes & Yes & 1000 & 8 & Novice, Intermediate, Expert & Full-time creator \\
P4 & No & Yes & Yes & 300 & 20 & Novice, Intermediate & Corporate trainer \\
P5 & No & Yes & Yes & 2000 & 30 & Novice, Intermediate, Expert &  Full-time creator \\
P6 & Yes & Yes & No & 50 & 2 & Intermediate, Expert & Excel consultant \\
P7 & Yes & Yes & Yes & 300 & 10 & Novice, Intermediate, Expert & Civil engineer \\
P8 & Yes & Yes & Yes & 100 & 20 & Novice, Intermediate & Accountant \\
\bottomrule
\end{tabular}
\end{center}
\end{scriptsize}
\end{table*}

%% file: tbl_designsResults.tex
\begin{table*}

\caption{To what extent each design addressed gaps we found in the comment analysis. 
Each row denotes a gap found during our investigation, with a column for each design. Highlighted in yellow are gaps we had not considered as part of the initial design formation, but that creators determined could be potentially addressed with the design.
}
\label{tbl:designsResults}
\begin{footnotesize}
\begin{center}
\begin{tabular}{{@{}ccc@{}}}
\shortstack{Gap code} 
& \shortstack{Design 1 \\ Interactive video tutorials} 
& \shortstack{Design 2 \\ Identifying gaps, alternatives}
\\
\midrule
\multicolumn{1}{r}{\hl{Spreadsheet sharing}}
& \hl{Partially}
& Not
 \\
\multicolumn{1}{r}{Production issues} 
& Partially
& Partially
\\
\multicolumn{1}{r}{Missing steps} 
& Fully
& Partially
\\
\multicolumn{1}{r}{Learning journey} 
& Partially
& \hl{Partially}
\\
\multicolumn{1}{r}{Maintaining knowledge} 
& Partially
& Not
\\
\multicolumn{1}{r}{Unknown concepts} 
& Partially
& Not
\\
\multicolumn{1}{r}{\hl{Language issues}} 
& \hl{Partially}
& \hl{Partially}
\\
\multicolumn{1}{r}{Version issues} 
& Partially
& Fully
\\
\multicolumn{1}{r}{Configuration question} 
& Not
& Partially
\\
\\
\bottomrule
\end{tabular}
\end{center}
\end{footnotesize}
\end{table*}

%% file: tbl_videos.tex
\begin{table*}

\caption{The YouTube Excel tutorial videos we used in our analysis. `yt=' is the video's YouTube ID.}
\label{tbl:videos}
\begin{tiny}
\begin{center}
\begin{tabular}{@{}ll@{}}
ID 
& Title \\
\midrule
V01 
& Microsoft Excel Tutorial - Beginners Level 1 { (yt=UrZZoDw9Yfg)} \\
V02 
& Excel Tutorial for Beginners | Excel Made Easy { (yt=0tdlR1rBwkM)} \\
V03 
& Microsoft Excel Tutorial for Beginners - Full Course { (yt=Vl0H-qTclOg)} \\
V04 
& The Beginner's Guide to Excel - Excel Basics Tutorial { (yt=rwbho0CgEAE)} \\
V05 
& The Ultimate Excel Tutorial - Beginner to Advanced - 5 Hours!
{ (yt=TpOIGij43AA)} \\
V06 
& Excel Tutorial | Microsoft Excel Tutorial | Excel Training | Intellipaat
 { (yt=27dxBp0EgCc)} \\
V07 
& VLOOKUP Tutorial for Excel - Everything You Need To Know {(yt=d3BYVQ6xIE4)} \\
V08 
& 3 Essential Excel skills for the data analyst { (yt=I1XeDS-GLbg)} \\
V09 
& Excel Formulas and Functions Tutorial { (yt=Jl0Qk63z2ZY)} \\
V10 
& Intermediate Excel Skills, Tips, and Tricks Tutorial { (yt=lxq\_46nY43g)} \\
V11 
& Pivot Table Excel Tutorial { (yt=m0wI61ahfLc)} \\
V12 
& Beginner to Pro FREE Excel Data Analysis Course { (yt=v2oNWja7M2E)} \\
V13 
& How to use Microsoft Excel - Beginner to Intermediate Class (with sample files){ (yt=F7aPazuS8QY)} \\
V14 
& Introduction to Microsoft Excel - Excel Basics Tutorial { (yt=fcbB0nkDik8)} \\
V15 
& How to Pass Excel Assessment Test For Job Applications - Step by Step Tutorial... { (yt=b-GxQvV9SWg)} \\
V16 
& A Real-Life Excel Test from a Job Interview: Can You Pass?? { (yt=7DcXo4q9tbY)} \\
V17 
& 50 Ultimate Excel Tips and Tricks for 2020 { (yt=FXs3WG7M-qk)} \\
V18 
& Basic Excel Formulas and Functions You NEED to KNOW! { (yt=y1126PQ5zRU)} \\
V19 
& Top 10 Most Important Excel Formulas - Made Easy! { (yt=ShBTJrdioLo)} \\
V20 
& Excel 2019 Beginner Tutorial { (yt=6JnEYGxxd8w)} \\
V21 
& Microsoft Excel Tutorial - Beginners Level 4 { (yt=c8qePWuYleg)} \\
V22 
& Excel 2019 Advanced Tutorial { (yt=bezV5U0dlbo)} \\
V23 
& VLOOKUP in Excel | Tutorial for Beginners { (yt=DZEPA9UhLBw)} \\
V24 
& Microsoft Excel Tutorial for Beginners | Excel Training | FREE Online Excel course { (yt=ormRboQsB-I)} \\
V25 
& Microsoft Excel Tutorial for Beginners | Excel Training | Excel Formulas and Functions... { (yt=RdTozKPY\_OQ)} \\
V26 
& Excel 2021 Advanced Tutorial { (yt=y4bA2CNScg8)} \\
V27 
& Excel Tutorial: Learn Excel in 30 Minutes - Just Right for your New Job Application { (yt=7RCdzTpKO0A)} \\
V28 
& MS Excel - Pivot Table Example 1 Video Tutorials { (yt=4PWVFBiFVVU)} \\
V29 
& How to build Interactive Excel Dashboards that Update with ONE CLICK! { (yt=K74\_FNnlIF8)} \\
V30 
& Learn Microsoft Excel Tutorial For Beginners in UNDER 45 MINUTES!... { (yt=djzmmeQM-ew)} \\
\bottomrule
\end{tabular}
\end{center}
\end{tiny}
\end{table*}

%% file: tbl_ttvideos.tex
\begin{table*}

\caption{The TikTok Excel tutorial videos we used in our analysis. `tt=' is the video's TikTok ID.}
\label{tbl:ttvideos}
\begin{tiny}
\begin{center}
\begin{tabular}{@{}ll@{}}
ID 
& Title \\
\midrule
V31 
& 3 Excel hacks all beginners should know! { (tt=7133215260080885038)} \\
V32 
& EXCEL TUTORIAL: Listen up Excel Beginners! { (tt=7143682564853665029)} \\
V33 
& Thousands of entries in seconds { (tt=7116501804191632686)} \\
V34 
& Am I the best co-worker of the year? { (tt=7025956874830761221)} \\
V35 
& If you found this useful, share this with a friend and follow for more hacks { (tt=7123623406100090117)} \\
V36 
& 5 Logical Formulas Beginners Should Know! { (tt=7090570128412167466)} \\
V37 
& Three Excel features to make your life easier! Who knew all three? { (tt=7089130412701764907)} \\
V38 
& (no title) { (tt=7033834780034747654)} \\
V39 
& My co-worker spent the rest of the day napping { (tt=7070131372345543979)} \\
V40 
& Do you need to learn VLOOKUP for work? { (tt=7041186723325316357)} \\
V41 
& Data Entry Clerks: this tip was made for you! { (tt=7065307256488807727)} \\
V42 
& (no title) { (tt=7054548921300831489)} \\
V43 
& What mistakes have you made in Excel? { (tt=7143961875758239022)} \\
V44 
& Don’t make this common Excel mistake! { (tt=7119902439872007470)} \\
V45 & 
The shortcut we have all been dreaming of { (tt=7129929626763250986)} \\
V46 & 
Excel Sequence { (tt=7101959854754925826)} \\
V47 & 
Replying to @renghahenish This is how to create a simple macro in 1 min { (tt=7154065532436712710)} \\
V48 & 
How to use alt = in Excel { (tt=7132844391982484778)} \\
V49 & 
Learn Excel: Join Tables Like a Pro { (tt=7142936116033834286)} \\
V50 & 
Calculating hours worked
 { (tt=7157747240335985921)} \\
V51 
& (no title) { (tt=7061691310230572294)} \\
V52 
& Don't forget to use this code { (tt=7116028520572194053)} \\
V53 
& Learn Excel: Data Prep with Power Query { (tt=7129550248997014830)} \\
V54 
& \#Tutorial on how easily to Auto Sum in Excel { (tt=7021374948509830426)} \\
V55 
& How to compare lists in Excel. { (tt=7159514315207232811)} \\
V56 
& How to fix \#REF! Error! { (tt=7140266589693381934)} \\
V57 
& Just download Libre Barcode 39 from Google to create barcodes in Excel
 { (tt=7159186239126523178)} \\
V58 
& When LEFT meets FIND = perfect duo { (tt=6989961642196995333)} \\
V59 
& Learn Excel: Pivot Table From Multiple Tab { (tt=7147076130808450350)} \\
V60 
& Please don’t tell my boss { (tt=7074568870262017322)} \\
\bottomrule
\end{tabular}
\end{center}
\end{tiny}
\end{table*}

%% file: tbl_instavideos.tex
\begin{table*}

\caption{The Instagram Excel tutorial videos we used in our analysis. `ig=' is the video's Instagram ID.}
\label{tbl:igvideos}
\begin{tiny}
\begin{center}
\begin{tabular}{@{}ll@{}}
ID 
& Title \\
\midrule
V61 
& EXCEL STOP Merging Cells { (ig=CkbPURDOW1a)} \\
V62 
& EXCEL VSTACK \& HSTACK { (ig=CkwDB\_TOlpd)} \\
V63 
& EXCEL Expiry Alert { (ig=CiabTk1sXcd)} \\
V64 
& Don’t type days manually! { (ig=ClOjQ3Aj\_nx)} \\
V65 
& Delete blank rows in Excel { (ig=ClBtRDvjQ6Z)} \\
V66 
& Number 1000 Cells in seconds in Excel! { (ig=CkbEyBBjRwr)} \\
V67 
& Learn about planets in Excel { (ig=CkgrlAjISJ5)} \\
V68 
& How to create barcodes in Excel { (ig=CkOt-KEoILG)} \\
V69 
& The VSTACK function allows you to stack sections of data on top of each other { (ig=Cj45dlfoVxM)} \\
V70 
& How to track your stocks in Excel { (ig=ClCfbjmJQm-)} \\
V71 
& Filter data with the click of a button using Slicers { (ig=Ck9QLwqJhiF)} \\
V72 
& Tired of seeing the value of zero? Just hide them using this tip { (ig=ClHYXCspUI6)} \\
V73 
& X lookup { (ig=CkQcdw9gcRt)} \\
V74 
& Do follow (Excel shortcuts) { (ig=CakVVz-AeMB)} \\
V75 
& Do follow (AutoSum using go to window) { (ig=CacanQZJb3u)} \\
V76 
& Use If function in Excel { (ig=CjCM-EHpPxS)} \\
V77 
& Calculate your age in Years, Months and Days in Excel { (ig=CjJ9e3rODi7)} \\
V78 
& How to split text into columns { (ig=Cj7lP8pL6PQ)} \\
V79 
& Ready to split your text in seconds?! { (ig=Cj-a3E1KRWH)} \\
V80 
& This excel tip is a game changer Say goodbye to messy data and hello to one unique list! { (ig=CkjIaj8Dy4M)} \\
V81 
& Combine data quickly with the VSTACK hack! { (ig=CkQpABtKb8X)} \\
V82 
& How to combine multiple sheets in Excel? { (ig=CkYW5VQvqGe)} \\
V83 
& How to convert a PDF into an Excel file? { (ig=CkLexVeMqWC)} \\
V84 
& How to make a large number list in Excel { (ig=Chzj470vMvx)} \\
V85 
& Create your 2023 calendar in seconds { (ig=ClggEFTA39F)} \\
V86 
& Importing location data is one of my forever favorite Excel hacks { (ig=ClTldDwgMnU)} \\
V87 
& Take filtering to the next level with slicers { (ig=ClBlAR9AcbJ)} \\
V88 
& Alright, time to settle this ONCE AND FOR ALL { (ig=CgbqI9AgCjB)} \\
V89 
& Now, should I inform my boss this Excel hack?? { (ig=CbKRPnag2xo)} \\
V90 
& POV: it’s the 21st century, time to STOP copying and pasting data { (ig=ChCkrVngr\_l)} \\
\bottomrule
\end{tabular}
\end{center}
\end{tiny}
\end{table*}

%% file: main.bbl
\begin{thebibliography}{10}
\expandafter\ifx\csname url\endcsname\relax
  \def\url#1{\texttt{#1}}\fi
\expandafter\ifx\csname urlprefix\endcsname\relax\def\urlprefix{URL }\fi
\expandafter\ifx\csname href\endcsname\relax
  \def\href#1#2{#2} \def\path#1{#1}\fi

\bibitem{kim2014crowdsourcing}
J.~Kim, P.~T. Nguyen, S.~Weir, P.~J. Guo, R.~C. Miller, K.~Z. Gajos,
  \href{https://doi.org/10.1145/2556288.2556986}{Crowdsourcing step-by-step
  information extraction to enhance existing how-to videos}, in: Proceedings of
  the SIGCHI Conference on Human Factors in Computing Systems, CHI '14,
  Association for Computing Machinery, New York, NY, USA, 2014, p. 4017–4026.
\newblock \href {https://doi.org/10.1145/2556288.2556986}
  {\path{doi:10.1145/2556288.2556986}}.
\newline\urlprefix\url{https://doi.org/10.1145/2556288.2556986}

\bibitem{kiani2019beyond}
K.~Kiani, G.~Cui, A.~Bunt, J.~McGrenere, P.~K. Chilana,
  \href{https://doi.org/10.1145/3290605.3300570}{Beyond "one-size-fits-all":
  Understanding the diversity in how software newcomers discover and make use
  of help resources}, in: Proceedings of the 2019 CHI Conference on Human
  Factors in Computing Systems, CHI '19, Association for Computing Machinery,
  New York, NY, USA, 2019, p. 1–14.
\newblock \href {https://doi.org/10.1145/3290605.3300570}
  {\path{doi:10.1145/3290605.3300570}}.
\newline\urlprefix\url{https://doi.org/10.1145/3290605.3300570}

\bibitem{end-user-encounters-with-lambda-22}
A.~Sarkar, S.~S. Ragavan, J.~Williams, A.~D. Gordon, End-user encounters with
  lambda abstraction in spreadsheets: Apollo’s bow or achilles’ heel?, in:
  2022 IEEE Symposium on Visual Languages and Human-Centric Computing (VL/HCC),
  {IEEE} Computer Society, Roma, Italy, 2022, pp. 1--11.
\newblock \href {https://doi.org/10.1109/VL/HCC53370.2022.9833131}
  {\path{doi:10.1109/VL/HCC53370.2022.9833131}}.

\bibitem{DBLP:conf/wcre/RoyHAWD16}
S.~Roy, F.~Hermans, E.~Aivaloglou, J.~Winter, A.~van Deursen, Evaluating
  automatic spreadsheet metadata extraction on a large set of responses from
  mooc participants, in: 2016 IEEE 23rd International Conference on Software
  Analysis, Evolution, and Reengineering (SANER), Vol.~1, {IEEE} Computer
  Society, Osaka, Japan, 2016, pp. 135--145.
\newblock \href {https://doi.org/10.1109/SANER.2016.98}
  {\path{doi:10.1109/SANER.2016.98}}.

\bibitem{sarkar2018spreadsheetlearning}
A.~Sarkar, A.~D. Gordon,
  \href{https://ppig.org/papers/2018-ppig-29th-sarkar/}{How do people learn to
  use spreadsheets? (work in progress)}, in: Proceedings of the 29th Annual
  Workshop of the Psychology of Programming Interest Group, {PPIG} 2018,
  London, UK, September 5 - 7, 2018, Psychology of Programming Interest Group,
  London, UK, 2018, pp. 28--35.
\newline\urlprefix\url{https://ppig.org/papers/2018-ppig-29th-sarkar/}

\bibitem{datareport}
Kepios, Global social media statistics,
  \url{https://datareportal.com/social-media-users/}, accessed: 2023-02-01
  (2023).

\bibitem{sarkar2023should}
A.~Sarkar, \href{https://doi.org/10.1145/3544549.3582741}{Should computers be
  easy to use? questioning the doctrine of simplicity in user interface
  design}, in: Extended Abstracts of the 2023 CHI Conference on Human Factors
  in Computing Systems, CHI EA '23, Association for Computing Machinery, New
  York, NY, USA, 2023.
\newblock \href {https://doi.org/10.1145/3544549.3582741}
  {\path{doi:10.1145/3544549.3582741}}.
\newline\urlprefix\url{https://doi.org/10.1145/3544549.3582741}

\bibitem{ko2004six}
A.~J. Ko, B.~A. Myers, H.~H. Aung,
  \href{https://doi.org/10.1109/VLHCC.2004.47}{Six learning barriers in
  end-user programming systems}, in: Proceedings of the 2004 IEEE Symposium on
  Visual Languages - Human Centric Computing, VLHCC '04, IEEE Computer Society,
  USA, 2004, p. 199–206.
\newblock \href {https://doi.org/10.1109/VLHCC.2004.47}
  {\path{doi:10.1109/VLHCC.2004.47}}.
\newline\urlprefix\url{https://doi.org/10.1109/VLHCC.2004.47}

\bibitem{potthast2021dilemma}
M.~Potthast, M.~Hagen, B.~Stein,
  \href{https://doi.org/10.1145/3451964.3451978}{The dilemma of the direct
  answer}, SIGIR Forum 54~(1) (Feb 2021).
\newblock \href {https://doi.org/10.1145/3451964.3451978}
  {\path{doi:10.1145/3451964.3451978}}.
\newline\urlprefix\url{https://doi.org/10.1145/3451964.3451978}

\bibitem{sarkar2016constructivist}
A.~Sarkar, \href{https://doi.org/10.1145/2851581.2892547}{Constructivist design
  for interactive machine learning}, in: Proceedings of the 2016 CHI Conference
  Extended Abstracts on Human Factors in Computing Systems, CHI EA '16,
  Association for Computing Machinery, New York, NY, USA, 2016, p. 1467–1475.
\newblock \href {https://doi.org/10.1145/2851581.2892547}
  {\path{doi:10.1145/2851581.2892547}}.
\newline\urlprefix\url{https://doi.org/10.1145/2851581.2892547}

\bibitem{srinivasa2016foraging}
S.~Srinivasa~Ragavan, S.~K. Kuttal, C.~Hill, A.~Sarma, D.~Piorkowski,
  M.~Burnett, \href{https://doi.org/10.1145/2858036.2858469}{Foraging among an
  overabundance of similar variants}, in: Proceedings of the 2016 CHI
  Conference on Human Factors in Computing Systems, CHI '16, Association for
  Computing Machinery, New York, NY, USA, 2016, p. 3509–3521.
\newblock \href {https://doi.org/10.1145/2858036.2858469}
  {\path{doi:10.1145/2858036.2858469}}.
\newline\urlprefix\url{https://doi.org/10.1145/2858036.2858469}

\bibitem{grossman2009survey}
T.~Grossman, G.~Fitzmaurice, R.~Attar,
  \href{https://doi.org/10.1145/1518701.1518803}{A survey of software
  learnability: Metrics, methodologies and guidelines}, in: Proceedings of the
  SIGCHI Conference on Human Factors in Computing Systems, CHI '09, Association
  for Computing Machinery, New York, NY, USA, 2009, p. 649–658.
\newblock \href {https://doi.org/10.1145/1518701.1518803}
  {\path{doi:10.1145/1518701.1518803}}.
\newline\urlprefix\url{https://doi.org/10.1145/1518701.1518803}

\bibitem{harrison1995comparison}
S.~M. Harrison, \href{https://doi.org/10.1145/223904.223915}{A comparison of
  still, animated, or nonillustrated on-line help with written or spoken
  instructions in a graphical user interface}, in: Proceedings of the SIGCHI
  Conference on Human Factors in Computing Systems, CHI '95, ACM
  Press/Addison-Wesley Publishing Co., USA, 1995, p. 82–89.
\newblock \href {https://doi.org/10.1145/223904.223915}
  {\path{doi:10.1145/223904.223915}}.
\newline\urlprefix\url{https://doi.org/10.1145/223904.223915}

\bibitem{chi2012mixt}
P.-Y. Chi, S.~Ahn, A.~Ren, M.~Dontcheva, W.~Li, B.~Hartmann,
  \href{https://doi.org/10.1145/2380116.2380130}{Mixt: Automatic generation of
  step-by-step mixed media tutorials}, in: Proceedings of the 25th Annual ACM
  Symposium on User Interface Software and Technology, UIST '12, Association
  for Computing Machinery, New York, NY, USA, 2012, p. 93–102.
\newblock \href {https://doi.org/10.1145/2380116.2380130}
  {\path{doi:10.1145/2380116.2380130}}.
\newline\urlprefix\url{https://doi.org/10.1145/2380116.2380130}

\bibitem{fourney2011query}
A.~Fourney, R.~Mann, M.~Terry,
  \href{https://doi.org/10.1145/2047196.2047224}{Query-feature graphs: Bridging
  user vocabulary and system functionality}, in: Proceedings of the 24th Annual
  ACM Symposium on User Interface Software and Technology, UIST '11,
  Association for Computing Machinery, New York, NY, USA, 2011, p. 207–216.
\newblock \href {https://doi.org/10.1145/2047196.2047224}
  {\path{doi:10.1145/2047196.2047224}}.
\newline\urlprefix\url{https://doi.org/10.1145/2047196.2047224}

\bibitem{fraser2020remap}
C.~A. Fraser, J.~M. Markel, N.~J. Basa, M.~Dontcheva, S.~Klemmer,
  \href{https://doi.org/10.1145/3379337.3415592}{Remap: Lowering the barrier to
  help-seeking with multimodal search}, in: Proceedings of the 33rd Annual ACM
  Symposium on User Interface Software and Technology, UIST '20, Association
  for Computing Machinery, New York, NY, USA, 2020, p. 979–986.
\newblock \href {https://doi.org/10.1145/3379337.3415592}
  {\path{doi:10.1145/3379337.3415592}}.
\newline\urlprefix\url{https://doi.org/10.1145/3379337.3415592}

\bibitem{fraser2019replay}
C.~A. Fraser, T.~J. Ngoon, M.~Dontcheva, S.~Klemmer,
  \href{https://doi.org/10.1145/3290605.3300527}{Replay: Contextually
  presenting learning videos across software applications}, in: Proceedings of
  the 2019 CHI Conference on Human Factors in Computing Systems, CHI '19,
  Association for Computing Machinery, New York, NY, USA, 2019, p. 1–13.
\newblock \href {https://doi.org/10.1145/3290605.3300527}
  {\path{doi:10.1145/3290605.3300527}}.
\newline\urlprefix\url{https://doi.org/10.1145/3290605.3300527}

\bibitem{bergman2005docwizards}
L.~Bergman, V.~Castelli, T.~Lau, D.~Oblinger,
  \href{https://doi.org/10.1145/1095034.1095067}{Docwizards: A system for
  authoring follow-me documentation wizards}, in: Proceedings of the 18th
  Annual ACM Symposium on User Interface Software and Technology, UIST '05,
  Association for Computing Machinery, New York, NY, USA, 2005, p. 191–200.
\newblock \href {https://doi.org/10.1145/1095034.1095067}
  {\path{doi:10.1145/1095034.1095067}}.
\newline\urlprefix\url{https://doi.org/10.1145/1095034.1095067}

\bibitem{leshed2008coscripter}
G.~Leshed, E.~M. Haber, T.~Matthews, T.~Lau,
  \href{https://doi.org/10.1145/1357054.1357323}{Coscripter: Automating \&
  sharing how-to knowledge in the enterprise}, in: Proceedings of the SIGCHI
  Conference on Human Factors in Computing Systems, CHI '08, Association for
  Computing Machinery, New York, NY, USA, 2008, p. 1719–1728.
\newblock \href {https://doi.org/10.1145/1357054.1357323}
  {\path{doi:10.1145/1357054.1357323}}.
\newline\urlprefix\url{https://doi.org/10.1145/1357054.1357323}

\bibitem{grabler2009generating}
F.~Grabler, M.~Agrawala, W.~Li, M.~Dontcheva, T.~Igarashi,
  \href{https://doi.org/10.1145/1531326.1531372}{Generating photo manipulation
  tutorials by demonstration}, ACM Trans. Graph. 28~(3) (jul 2009).
\newblock \href {https://doi.org/10.1145/1531326.1531372}
  {\path{doi:10.1145/1531326.1531372}}.
\newline\urlprefix\url{https://doi.org/10.1145/1531326.1531372}

\bibitem{lafreniere2014investigating}
B.~Lafreniere, T.~Grossman, J.~Matejka, G.~Fitzmaurice,
  \href{https://doi.org/10.1145/2556288.2557142}{Investigating the feasibility
  of extracting tool demonstrations from in-situ video content}, in:
  Proceedings of the SIGCHI Conference on Human Factors in Computing Systems,
  CHI '14, Association for Computing Machinery, New York, NY, USA, 2014, p.
  4007–4016.
\newblock \href {https://doi.org/10.1145/2556288.2557142}
  {\path{doi:10.1145/2556288.2557142}}.
\newline\urlprefix\url{https://doi.org/10.1145/2556288.2557142}

\bibitem{whitehill2017crowdsourcing}
J.~Whitehill, M.~Seltzer, \href{https://doi.org/10.1145/3051457.3053973}{A
  crowdsourcing approach to collecting tutorial videos -- toward personalized
  learning-at-scale}, in: Proceedings of the Fourth (2017) ACM Conference on
  Learning @ Scale, L@S '17, Association for Computing Machinery, New York, NY,
  USA, 2017, p. 157–160.
\newblock \href {https://doi.org/10.1145/3051457.3053973}
  {\path{doi:10.1145/3051457.3053973}}.
\newline\urlprefix\url{https://doi.org/10.1145/3051457.3053973}

\bibitem{guo2014video}
P.~J. Guo, J.~Kim, R.~Rubin, \href{https://doi.org/10.1145/2556325.2566239}{How
  video production affects student engagement: An empirical study of mooc
  videos}, in: Proceedings of the First ACM Conference on Learning @ Scale
  Conference, L@S '14, Association for Computing Machinery, New York, NY, USA,
  2014, p. 41–50.
\newblock \href {https://doi.org/10.1145/2556325.2566239}
  {\path{doi:10.1145/2556325.2566239}}.
\newline\urlprefix\url{https://doi.org/10.1145/2556325.2566239}

\bibitem{pongnumkul2011pause}
S.~Pongnumkul, M.~Dontcheva, W.~Li, J.~Wang, L.~Bourdev, S.~Avidan, M.~F.
  Cohen, \href{https://doi.org/10.1145/2047196.2047213}{Pause-and-play:
  Automatically linking screencast video tutorials with applications}, in:
  Proceedings of the 24th Annual ACM Symposium on User Interface Software and
  Technology, UIST '11, Association for Computing Machinery, New York, NY, USA,
  2011, p. 135–144.
\newblock \href {https://doi.org/10.1145/2047196.2047213}
  {\path{doi:10.1145/2047196.2047213}}.
\newline\urlprefix\url{https://doi.org/10.1145/2047196.2047213}

\bibitem{pavel2014video}
A.~Pavel, C.~Reed, B.~Hartmann, M.~Agrawala,
  \href{https://doi.org/10.1145/2642918.2647400}{Video digests: A browsable,
  skimmable format for informational lecture videos}, in: Proceedings of the
  27th Annual ACM Symposium on User Interface Software and Technology, UIST
  '14, Association for Computing Machinery, New York, NY, USA, 2014, p.
  573–582.
\newblock \href {https://doi.org/10.1145/2642918.2647400}
  {\path{doi:10.1145/2642918.2647400}}.
\newline\urlprefix\url{https://doi.org/10.1145/2642918.2647400}

\bibitem{weir2015learnersourcing}
S.~Weir, J.~Kim, K.~Z. Gajos, R.~C. Miller,
  \href{https://doi.org/10.1145/2675133.2675219}{Learnersourcing subgoal labels
  for how-to videos}, in: Proceedings of the 18th ACM Conference on Computer
  Supported Cooperative Work \& Social Computing, CSCW '15, Association for
  Computing Machinery, New York, NY, USA, 2015, p. 405–416.
\newblock \href {https://doi.org/10.1145/2675133.2675219}
  {\path{doi:10.1145/2675133.2675219}}.
\newline\urlprefix\url{https://doi.org/10.1145/2675133.2675219}

\bibitem{kim2014data}
J.~Kim, P.~J. Guo, C.~J. Cai, S.-W.~D. Li, K.~Z. Gajos, R.~C. Miller,
  \href{https://doi.org/10.1145/2642918.2647389}{Data-driven interaction
  techniques for improving navigation of educational videos}, in: Proceedings
  of the 27th Annual ACM Symposium on User Interface Software and Technology,
  UIST '14, Association for Computing Machinery, New York, NY, USA, 2014, p.
  563–572.
\newblock \href {https://doi.org/10.1145/2642918.2647389}
  {\path{doi:10.1145/2642918.2647389}}.
\newline\urlprefix\url{https://doi.org/10.1145/2642918.2647389}

\bibitem{kim2014understanding}
J.~Kim, P.~J. Guo, D.~T. Seaton, P.~Mitros, K.~Z. Gajos, R.~C. Miller,
  \href{https://doi.org/10.1145/2556325.2566237}{Understanding in-video
  dropouts and interaction peaks inonline lecture videos}, in: Proceedings of
  the First ACM Conference on Learning @ Scale Conference, L@S '14, Association
  for Computing Machinery, New York, NY, USA, 2014, p. 31–40.
\newblock \href {https://doi.org/10.1145/2556325.2566237}
  {\path{doi:10.1145/2556325.2566237}}.
\newline\urlprefix\url{https://doi.org/10.1145/2556325.2566237}

\bibitem{rettig1991nobody}
M.~Rettig, Nobody reads documentation, Communications of the ACM 34~(7) (1991)
  19--24.

\bibitem{rieman1996field}
J.~Rieman, A field study of exploratory learning strategies, ACM Transactions
  on Computer-Human Interaction (TOCHI) 3~(3) (1996) 189--218.

\bibitem{lazar2006workplace}
J.~Lazar, A.~Jones, B.~Shneiderman, Workplace user frustration with computers:
  An exploratory investigation of the causes and severity, Behaviour \&
  Information Technology 25~(03) (2006) 239--251.

\bibitem{carroll1987paradox}
J.~M. Carroll, M.~B. Rosson, Paradox of the Active User, MIT Press, Cambridge,
  MA, USA, 1987, Ch.~5, p. 80–111.

\bibitem{farkas1993role}
D.~K. Farkas, The role of balloon help, ACM SIGDOC Asterisk Journal of Computer
  Documentation 17~(2) (1993) 3--19.

\bibitem{huang2007graphstract}
J.~Huang, M.~B. Twidale,
  \href{https://doi.org/10.1145/1294211.1294248}{Graphstract: Minimal graphical
  help for computers}, in: Proceedings of the 20th Annual ACM Symposium on User
  Interface Software and Technology, UIST '07, Association for Computing
  Machinery, New York, NY, USA, 2007, p. 203–212.
\newblock \href {https://doi.org/10.1145/1294211.1294248}
  {\path{doi:10.1145/1294211.1294248}}.
\newline\urlprefix\url{https://doi.org/10.1145/1294211.1294248}

\bibitem{lau2004sheepdog}
T.~Lau, L.~Bergman, V.~Castelli, D.~Oblinger,
  \href{https://doi.org/10.1145/964442.964464}{Sheepdog: Learning procedures
  for technical support}, in: Proceedings of the 9th International Conference
  on Intelligent User Interfaces, IUI '04, Association for Computing Machinery,
  New York, NY, USA, 2004, p. 109–116.
\newblock \href {https://doi.org/10.1145/964442.964464}
  {\path{doi:10.1145/964442.964464}}.
\newline\urlprefix\url{https://doi.org/10.1145/964442.964464}

\bibitem{kelleher2005stencils}
C.~Kelleher, R.~Pausch,
  \href{https://doi.org/10.1145/1054972.1055047}{Stencils-based tutorials:
  Design and evaluation}, in: Proceedings of the SIGCHI Conference on Human
  Factors in Computing Systems, CHI '05, Association for Computing Machinery,
  New York, NY, USA, 2005, p. 541–550.
\newblock \href {https://doi.org/10.1145/1054972.1055047}
  {\path{doi:10.1145/1054972.1055047}}.
\newline\urlprefix\url{https://doi.org/10.1145/1054972.1055047}

\bibitem{carroll1984training}
J.~M. Carroll, C.~Carrithers, Training wheels in a user interface,
  Communications of the ACM 27~(8) (1984) 800--806.

\bibitem{mcgrenere2002evaluation}
J.~McGrenere, R.~M. Baecker, K.~S. Booth,
  \href{https://doi.org/10.1145/503376.503406}{An evaluation of a multiple
  interface design solution for bloated software}, in: Proceedings of the
  SIGCHI Conference on Human Factors in Computing Systems, CHI '02, Association
  for Computing Machinery, New York, NY, USA, 2002, p. 164–170.
\newblock \href {https://doi.org/10.1145/503376.503406}
  {\path{doi:10.1145/503376.503406}}.
\newline\urlprefix\url{https://doi.org/10.1145/503376.503406}

\bibitem{mcgrenere2007field}
J.~McGrenere, R.~M. Baecker, K.~S. Booth, A field evaluation of an adaptable
  two-interface design for feature-rich software, ACM Transactions on
  Computer-Human Interaction (TOCHI) 14~(1) (2007) 3--es.

\bibitem{lafreniere2014task}
B.~Lafreniere, A.~Bunt, M.~Terry,
  \href{https://doi.org/10.1145/2686612.2686620}{Task-centric interfaces for
  feature-rich software}, in: Proceedings of the 26th Australian Computer-Human
  Interaction Conference on Designing Futures: The Future of Design, OzCHI '14,
  Association for Computing Machinery, New York, NY, USA, 2014, p. 49–58.
\newblock \href {https://doi.org/10.1145/2686612.2686620}
  {\path{doi:10.1145/2686612.2686620}}.
\newline\urlprefix\url{https://doi.org/10.1145/2686612.2686620}

\bibitem{andrade2008expressing}
O.~D. Andrade, D.~G. Novick,
  \href{https://doi.org/10.1145/1456536.1456562}{Expressing help at appropriate
  levels}, in: Proceedings of the 26th Annual ACM International Conference on
  Design of Communication, SIGDOC '08, Association for Computing Machinery, New
  York, NY, USA, 2008, p. 125–130.
\newblock \href {https://doi.org/10.1145/1456536.1456562}
  {\path{doi:10.1145/1456536.1456562}}.
\newline\urlprefix\url{https://doi.org/10.1145/1456536.1456562}

\bibitem{van2006paradox}
C.~C. van Nimwegen, D.~Burgos, H.~H. van Oostendorp, H.~H. J.~M. Schijf,
  \href{https://doi.org/10.1145/1124772.1124908}{The paradox of the assisted
  user: Guidance can be counterproductive}, in: Proceedings of the SIGCHI
  Conference on Human Factors in Computing Systems, CHI '06, Association for
  Computing Machinery, New York, NY, USA, 2006, p. 917–926.
\newblock \href {https://doi.org/10.1145/1124772.1124908}
  {\path{doi:10.1145/1124772.1124908}}.
\newline\urlprefix\url{https://doi.org/10.1145/1124772.1124908}

\bibitem{bar2019good}
S.~Bar-Or, J.~Meyer, What is good help? responses to solicited and unsolicited
  assistance, International Journal of Human--Computer Interaction 35~(2)
  (2019) 131--139.

\bibitem{xiao2004empirical}
J.~Xiao, J.~Stasko, R.~Catrambone, et~al., An empirical study of the effect of
  agent competence on user performance and perception, in: Proceedings of the
  Third International Joint Conference on Autonomous Agents and Multiagent
  Systems, 2004. AAMAS 2004., Vol.~4, IEEE, New York, NY, USA, 2004, pp.
  178--185.

\bibitem{grossman2007strategies}
T.~Grossman, P.~Dragicevic, R.~Balakrishnan,
  \href{https://doi.org/10.1145/1240624.1240865}{Strategies for accelerating
  on-line learning of hotkeys}, in: Proceedings of the SIGCHI Conference on
  Human Factors in Computing Systems, CHI '07, Association for Computing
  Machinery, New York, NY, USA, 2007, p. 1591–1600.
\newblock \href {https://doi.org/10.1145/1240624.1240865}
  {\path{doi:10.1145/1240624.1240865}}.
\newline\urlprefix\url{https://doi.org/10.1145/1240624.1240865}

\bibitem{barik2015heart}
T.~Barik, B.~Johnson, E.~Murphy-Hill,
  \href{https://doi.org/10.1145/2786805.2803200}{I heart hacker news: Expanding
  qualitative research findings by analyzing social news websites}, in:
  Proceedings of the 2015 10th Joint Meeting on Foundations of Software
  Engineering, ESEC/FSE 2015, Association for Computing Machinery, New York,
  NY, USA, 2015, p. 882–885.
\newblock \href {https://doi.org/10.1145/2786805.2803200}
  {\path{doi:10.1145/2786805.2803200}}.
\newline\urlprefix\url{https://doi.org/10.1145/2786805.2803200}

\bibitem{sarkar2022lambdas}
A.~Sarkar, S.~S. Ragavan, J.~Williams, A.~D. Gordon, End-user encounters with
  lambda abstraction in spreadsheets: Apollo’s bow or achilles’ heel?, in:
  2022 IEEE Symposium on Visual Languages and Human-Centric Computing (VL/HCC),
  IEEE Computer Society, Roma, Italy, 2022, pp. 1--11.
\newblock \href {https://doi.org/10.1109/VL/HCC53370.2022.9833131}
  {\path{doi:10.1109/VL/HCC53370.2022.9833131}}.

\bibitem{braun2006using}
V.~Braun, V.~Clarke, Using thematic analysis in psychology, Qualitative
  research in psychology 3~(2) (2006) 77--101.

\bibitem{campbell2013coding}
J.~L. Campbell, C.~Quincy, J.~Osserman, O.~K. Pedersen, Coding in-depth
  semistructured interviews: Problems of unitization and intercoder reliability
  and agreement, Sociological methods \& research 42~(3) (2013) 294--320.

\bibitem{mcdonald2019reliability}
N.~McDonald, S.~Schoenebeck, A.~Forte, Reliability and inter-rater reliability
  in qualitative research: Norms and guidelines for cscw and hci practice,
  Proceedings of the ACM on human-computer interaction 3~(CSCW) (2019) 1--23.

\bibitem{saldana2021coding}
J.~Salda{\~n}a, \href{https://books.google.co.uk/books?id=RwcVEAAAQBAJ}{The
  Coding Manual for Qualitative Researchers}, Core textbook, SAGE Publications,
  London, United Kingdom, 2021.
\newline\urlprefix\url{https://books.google.co.uk/books?id=RwcVEAAAQBAJ}

\bibitem{ragavan2021comprehension}
S.~Srinivasa~Ragavan, A.~Sarkar, A.~D. Gordon,
  \href{https://doi.org/10.1145/3411764.3445634}{Spreadsheet comprehension:
  Guesswork, giving up and going back to the author}, in: Proceedings of the
  2021 CHI Conference on Human Factors in Computing Systems, CHI '21,
  Association for Computing Machinery, New York, NY, USA, 2021.
\newblock \href {https://doi.org/10.1145/3411764.3445634}
  {\path{doi:10.1145/3411764.3445634}}.
\newline\urlprefix\url{https://doi.org/10.1145/3411764.3445634}

\bibitem{chalhoub2022freedom}
G.~Chalhoub, A.~Sarkar,
  \href{https://doi.org/10.1145/3491102.3501833}{“it’s freedom to put
  things where my mind wants”: Understanding and improving the user
  experience of structuring data in spreadsheets}, in: Proceedings of the 2022
  CHI Conference on Human Factors in Computing Systems, CHI '22, Association
  for Computing Machinery, New York, NY, USA, 2022.
\newblock \href {https://doi.org/10.1145/3491102.3501833}
  {\path{doi:10.1145/3491102.3501833}}.
\newline\urlprefix\url{https://doi.org/10.1145/3491102.3501833}

\bibitem{figma}
F.~Inc., Figma: the collaborative interface design tool,
  \url{https://www.figma.com/}, accessed: 2023-01-31 (2023).

\bibitem{srinivasa2021spreadsheet}
S.~Srinivasa~Ragavan, A.~Sarkar, A.~D. Gordon,
  \href{https://doi.org/10.1145/3411764.3445634}{Spreadsheet comprehension:
  Guesswork, giving up and going back to the author}, in: Proceedings of the
  2021 CHI Conference on Human Factors in Computing Systems, CHI '21,
  Association for Computing Machinery, New York, NY, USA, 2021.
\newblock \href {https://doi.org/10.1145/3411764.3445634}
  {\path{doi:10.1145/3411764.3445634}}.
\newline\urlprefix\url{https://doi.org/10.1145/3411764.3445634}

\bibitem{KindelDiscourse2017}
A.~Kindel, M.~Yeomans, J.~Reich, B.~Stewart, D.~Tingley,
  \href{https://doi.org/10.1145/3051457.3053967}{Discourse: Mooc discussion
  forum analysis at scale}, in: Proceedings of the Fourth (2017) ACM Conference
  on Learning @ Scale, L@S '17, Association for Computing Machinery, New York,
  NY, USA, 2017, p. 141–142.
\newblock \href {https://doi.org/10.1145/3051457.3053967}
  {\path{doi:10.1145/3051457.3053967}}.
\newline\urlprefix\url{https://doi.org/10.1145/3051457.3053967}

\bibitem{RUIPEREZVALIENTE2022104426}
J.~A. Ruipérez-Valiente, T.~Staubitz, M.~Jenner, S.~Halawa, J.~Zhang,
  I.~Despujol, J.~Maldonado-Mahauad, G.~Montoro, M.~Peffer, T.~Rohloff,
  J.~Lane, C.~Turro, X.~Li, M.~Pérez-Sanagustín, J.~Reich,
  \href{https://www.sciencedirect.com/science/article/pii/S0360131521003031}{Large
  scale analytics of global and regional mooc providers: Differences in
  learners’ demographics, preferences, and perceptions}, Computers \&
  Education 180 (2022) 104426.
\newblock \href {https://doi.org/https://doi.org/10.1016/j.compedu.2021.104426}
  {\path{doi:https://doi.org/10.1016/j.compedu.2021.104426}}.
\newline\urlprefix\url{https://www.sciencedirect.com/science/article/pii/S0360131521003031}

\bibitem{openai2023gpt4}
OpenAI, Gpt-4 technical report (2023).
\newblock \href {http://arxiv.org/abs/2303.08774} {\path{arXiv:2303.08774}}.

\bibitem{touvron2023llama}
H.~Touvron, L.~Martin, K.~Stone, P.~Albert, A.~Almahairi, Y.~Babaei,
  N.~Bashlykov, S.~Batra, P.~Bhargava, S.~Bhosale, D.~Bikel, L.~Blecher, C.~C.
  Ferrer, M.~Chen, G.~Cucurull, D.~Esiobu, J.~Fernandes, J.~Fu, W.~Fu,
  B.~Fuller, C.~Gao, V.~Goswami, N.~Goyal, A.~Hartshorn, S.~Hosseini, R.~Hou,
  H.~Inan, M.~Kardas, V.~Kerkez, M.~Khabsa, I.~Kloumann, A.~Korenev, P.~S.
  Koura, M.-A. Lachaux, T.~Lavril, J.~Lee, D.~Liskovich, Y.~Lu, Y.~Mao,
  X.~Martinet, T.~Mihaylov, P.~Mishra, I.~Molybog, Y.~Nie, A.~Poulton,
  J.~Reizenstein, R.~Rungta, K.~Saladi, A.~Schelten, R.~Silva, E.~M. Smith,
  R.~Subramanian, X.~E. Tan, B.~Tang, R.~Taylor, A.~Williams, J.~X. Kuan,
  P.~Xu, Z.~Yan, I.~Zarov, Y.~Zhang, A.~Fan, M.~Kambadur, S.~Narang,
  A.~Rodriguez, R.~Stojnic, S.~Edunov, T.~Scialom, Llama 2: Open foundation and
  fine-tuned chat models (2023).
\newblock \href {http://arxiv.org/abs/2307.09288} {\path{arXiv:2307.09288}}.

\bibitem{drosos2022}
I.~Drosos, P.~J. Guo, \href{https://doi.org/10.1145/3532106.3533489}{The design
  space of livestreaming equipment setups: Tradeoffs, challenges, and
  opportunities}, in: Designing Interactive Systems Conference, DIS '22,
  Association for Computing Machinery, New York, NY, USA, 2022, p. 835–848.
\newblock \href {https://doi.org/10.1145/3532106.3533489}
  {\path{doi:10.1145/3532106.3533489}}.
\newline\urlprefix\url{https://doi.org/10.1145/3532106.3533489}

\bibitem{drosos2021}
I.~Drosos, P.~J. Guo, Streamers teaching programming, art, and gaming:
  Cognitive apprenticeship, serendipitous teachable moments, and tacit expert
  knowledge, in: 2021 IEEE Symposium on Visual Languages and Human-Centric
  Computing (VL/HCC), IEEE, St Louis, MO, USA, 2021, pp. 1--6.
\newblock \href {https://doi.org/10.1109/VL/HCC51201.2021.9576481}
  {\path{doi:10.1109/VL/HCC51201.2021.9576481}}.

\bibitem{koskei_role_2015}
B.~Koskei, C.~Simiyu, Role of interviews, observation, pitfalls and ethical
  issues in qualitative research methods, Journal of Educational Policy and
  Entrepreneurial Research 2~(3) (2015) 108--117.

\bibitem{kajornboon_using_2005}
A.~B. Kajornboon, Using interviews as research instruments, E-journal for
  Research Teachers 2~(1) (2005) 1--9.

\bibitem{birmingham_using_2003}
P.~Birmingham, D.~Wilkinson, Using research instruments: {A} guide for
  researchers, Routledge, England, UK, 2003.

\bibitem{Jupp2006}
V.~Jupp, \href{https://doi.org/10.4135/9780857020116}{The {SAGE} Dictionary of
  Social Research Methods}, {SAGE} Publications, Ltd, London, UK, 2006.
\newblock \href {https://doi.org/10.4135/9780857020116}
  {\path{doi:10.4135/9780857020116}}.
\newline\urlprefix\url{https://doi.org/10.4135/9780857020116}

\bibitem{c_n_trueman_structured_2015}
{C N Trueman},
  \href{https://www.historylearningsite.co.uk/sociology/research-methods-in-sociology/structured-interviews/}{Structured
  {Interviews}} (May 2015).
\newline\urlprefix\url{https://www.historylearningsite.co.uk/sociology/research-methods-in-sociology/structured-interviews/}

\end{thebibliography}
